\documentclass[twoside,twocolumn,english,aps,prb,superscriptaddress]{revtex4-2}
\usepackage[T1]{fontenc}
\usepackage[latin9]{inputenc}
\usepackage{bm}
\usepackage{amsmath}
\usepackage{graphicx}

\makeatletter
\usepackage[colorlinks,linkcolor=blue,anchorcolor=blue,citecolor=red]{hyperref}

\makeatother

\usepackage{babel}
\begin{document}
\title{Many-body multipole indices revealed by the real-space dynamical mean-field
theory}
\author{Guoao Yang}
\affiliation{School of Physics and Optoelectronics Engineering, Anhui University,
Hefei, Anhui Province 230601, P.R. China }
\author{Jianhui Zhou}
\email{jhzhou@hmfl.ac.cn}

\affiliation{Anhui Key Laboratory of Low-energy Quantum Materials and Devices,
High Magnetic Field Laboratory, HFIPS, Chinese Academy of Sciences,
Hefei, Anhui 230031, China}
\author{Tao Qin}
\email{taoqin@ahu.edu.cn}

\affiliation{School of Physics and Optoelectronics Engineering, Anhui University,
Hefei, Anhui Province 230601, P.R. China }
\begin{abstract}
The multipole moments are fundamental properties of insulators, and
have attracted lots of attention with emerging of the higher-order
topological insulators. A couple of ways, including generalization
of the formula for the polarization and the Wilson loop, have been
proposed to calculate it in real materials. However, a practical method
to explore it in correlated insulators is still lacking. Here, we
proposed a systematic way, which combines the general Green's function
formula for multiopoles with the real-space dynamical mean-field theory,
to calculate the multipole moments in correlated materials. Our demonstrating
calculations are consistent with symmetry analysis, and the calculations
of the spectral functions further confirm our results. This method
opens the new avenue to study the topological phase transitions in
correlated multipole insulators and other crucial physical quantities
closely related to multipole moments.
\end{abstract}
\maketitle

\section{introduction}

The polarization is a fundamental property of insulators, and it has
been under intensive study for decades~\citep{resta_theory_1992,king-smith_theory_1993,resta_macroscopic_1994,ortiz_macroscopic_1994,coh_electric_2009,xiao_berry_2010,vaidya_polarization_2024}.
It is a ground-breaking progress to realize that the change of polarization
is closely related to the integral of the Berry curvature in the parameter
space~\citep{resta_theory_1992,king-smith_theory_1993}. The emergence
of topological insulators offers a new platform for the study of polarization,
which works as a topological index for the topological phase transition.
However, for a long time, a systematic method to reveal the polarization
in a correlated insulator remains elusive, even if there are detailed
discussion on this formalism~\citep{ortiz_macroscopic_1994,ortiz_quantum_1996}.
In Resta's pioneering work~\citep{resta_quantum-mechanical_1998},
a formula for the polarization is proposed, and affords the possibility
of exploring the polarization in the interacting case. Furthermore,
with the concept of localization length~\citep{resta_electron_1999},
its connection with Kohn's theory of insulating state~\citep{kohn_theory_1964}
has been established. To explore the properties of multipole moments,
which offers a distinctive point of view to uncover the nontrivial
topological physics, is fundamentally important.

The topics on the polarization and its multipole generalization have
become more active with the generalized concept of higher-order topological
insulator. The higher-order topological phase, which is a key playground
for multipole indices, is characterized by the quantized multipole
indices, pioneered by Benalcazar \emph{et al}.~\citep{benalcazar_quantized_2017}.
This novel concept has been identified in different non-crystalline
lattices without periodicity~\citep{xie_higher-order_2021,yang_higher-order_2024}.
The generalization of Resta's formula to the multipole case has  opened
the avenue to study the multipole~\citep{kang_many-body_2019}, which
can serve as an index to classify the higher-order topological insulators.
Even though a couple of issues have been pointed out in this generalization~\citep{ono_difficulties_2019},
detailed calculations have been done for non-interacting models~\citep{lee_many-body_2022},
and its effectiveness has been verified. Its relationship with the
Wilson-loop formulation~\citep{benalcazar_quantized_2017,benalcazar_electric_2017}
has been elaborated. Furthermore, several ways have been tried out
to study multipole indices of correlated insulator. In Ref.~\citep{nourafkan_electric_2013},
the polarization of the Hubbard model is explored with the dynamical
mean-field theory (DMFT)~\citep{georges_dynamical_1996}. However,
it is hard to apply it to the multipole moments. In Ref.~\citep{peng_correlation_2020},
correlation effects in quadrupole insulators are studied with the
quantum Monte Carlo simulations and topological Hamiltonians. In Ref.~\citep{kang_many-body_2021}
a many-body invariant is extracted from Resta's operator for the polarization
with methods of exact diagonalization and infinite density matrix
renormalization group. Clearly, these are indirect ways to calculate
the multipole indices in correlated systems. 

However, a practical procedure for directly calculating the many-body
multipole index of correlated insulators is still lacking. Here, we
propose a Green's function formalism in the real space for the multipole
indices, which can be perfectly fitted into the scheme of the real-space
dynamical mean-field theory (R-DMFT)~\citep{potthoff_self-energy-functional_2003,helmes_dynamical_2008,hofstetter_quantum_2018}.
We have carried out calculations of multipole indices of the correlated
Benalcazar-Bernevig-Hughes (BBH) model with different spatial symmetries.
The results are consistent with the symmetry analysis, and shed light
on the exploration of multipole indices of the correlated insulator.
Furthermore, we would like to point out that our formalism can be
adopted to study the higher-order topological phase in crystalline
systems with quenched disorder and in non-crystalline systems, and
that it also affords a way to compute the localization length of the
electron~\citep{resta_electron_1999}, the polarization fluctuation~\citep{resta_polarization_2006},
and the quantum geometric tensor~\citep{souza_polarization_2000}
in the correlated lattices. 

The rest of the manuscript has been organized as follows. In Sec.~\ref{sec:Analytical},
we show our derivation details for the general formula of the multipoles,
explain our protocol on how to combine them with the R-DMFT, and point
out its further generalization. In Sec.~\ref{sec:R-DMFT-calculation},
we show our demonstrating calculation results for the correlated BBH
model with different spatial symmetries. A summary and outlook is
presented in Sec.~\ref{sec:Outlook}. 

\section{\label{sec:Analytical}Many-body multipole indices in Green's function
formalism}

We propose a Green's function formalism for the electrical polarization
for a crystal, which is ready to be studied in a correlated system,
\begin{equation}
P_{x}=\frac{1}{2\pi\mathcal{S}}\mathrm{Im}\left[\mathrm{Tr}\ln\left(\tilde{G}_{x}+I\right)\right]-P_{bg}.\label{eq:Px}
\end{equation}
This is the central result in this work. Here $\tilde{G}_{x}=\alpha_{x}G$.
$\alpha_{x}=\mathrm{diag}\{e^{\frac{2\pi ix_{1}}{L_{x}}}-1,\cdots,e^{\frac{2\pi ix_{N}}{L_{x}}}-1\}$,
and $G_{x}$ is a matrix of Green's function with the $i,j$-th element
as $\left\langle c_{i}^{\dagger}c_{j}\right\rangle =\frac{1}{\beta}\sum_{n}G_{ij}\left(i\omega_{n}\right)$,
where $\omega_{n}=\frac{\left(2n+1\right)\pi}{\beta}$ is the Matsubara
frequency, and $c_{i}^{\dagger}$ ($c_{j}$) is the creation (annihilation)
operator on the lattice site $i$ ($j$) in a supercell with sizes
as $L_{x}$, $L_{x}L_{y}$, and $L_{x}L_{y}L_{z}$ for one, two and
three-dimensional cases, respectively. $I$ is the identity matrix.
$\mathcal{S}$ is dimension-dependent as $\mathcal{S}=1$, $L_{y}$,
$L_{y}L_{z}$ for different dimensions, which is the area perpendicular
to the $x$ direction. $P_{bg}=\frac{n_{f}}{\mathcal{S}}\sum_{i}\frac{x_{i}}{L_{x}}$
is the contribution from the background charge, with $n_{f}$ the
local density of electrical states. It can be combined with fixed
boundary condition, periodic boundary condition and a hybrid boundary
condition. The generalization to include the spin flavor is straightforward. 

We outline the crucial derivation steps for Eq.~(\ref{eq:Px}). Starting
from the formula~\citep{resta_quantum-mechanical_1998,kang_many-body_2019},
$P_{x}=\frac{1}{2\pi\mathcal{S}}\mathrm{Im}\left[\ln\left\langle \hat{U}_{x}\right\rangle \right]-P_{bg}$,
where $\left\langle \hat{U}_{x}\right\rangle =\left\langle GS\right|\hat{U}_{x}\left|GS\right\rangle =\left\langle GS\right|e^{2\pi i\sum_{l}\frac{x_{l}}{L_{x}}\hat{n}_{l}}\left|GS\right\rangle $,
we have, 
\begin{equation}
\left\langle \hat{U}_{x}\right\rangle =\left\langle GS\right|e^{2\pi i\frac{x_{1}}{L_{x}}\hat{n}_{1}}e^{2\pi i\frac{x_{2}}{L_{x}}\hat{n}_{2}}\cdots e^{2\pi i\frac{x_{N}}{L_{x}}\hat{n}_{N}}\left|GS\right\rangle .
\end{equation}
For the electron, we note the nilpotency property of the electron
number operator $\hat{n}_{l}^{2}=\hat{n}_{l}$, which is a key observation
here, and it follows that $e^{2\pi i\frac{x_{l}}{L_{x}}\hat{n}_{l}}=\sum_{m=0}^{\infty}\frac{1}{m!}\left(2\pi i\frac{x_{l}}{L_{x}}\hat{n}_{l}\right)^{m}=1+\alpha_{l}\hat{n}_{l}$
with $\alpha_{l}=e^{\frac{2\pi ix_{l}}{L_{x}}}-1$. Therefore, we
come to 
\begin{align}
\left\langle \hat{U}_{x}\right\rangle  & =\left\langle GS\right|\prod_{l=1}^{N}\left(1+\alpha_{l}\hat{n}_{l}\right)\left|GS\right\rangle ,\\
 & =1+\sum_{l=1}^{N}\alpha_{l}\left\langle \hat{n}_{l}\right\rangle +\sum_{1\le i<j\le N}\alpha_{i}\alpha_{j}\left\langle \hat{n}_{i}\hat{n}_{j}\right\rangle +\cdots,\label{eq:U_poly}
\end{align}
where the last step is obtained by a direct multiplication. We define
a $N\times N$ matrix $\tilde{G}_{x}$, with $\left(\tilde{G}_{x}\right)_{ij}=\alpha_{i}\left\langle c_{i}^{\dagger}c_{j}\right\rangle $,
($i,j=1,2,\cdots,N$). A second key observation~\citep{cheong_many-body_2004}
is that different orders of principal minors for $\det\tilde{G}_{x}$
coincides with different order of terms for $\left\langle \hat{U}_{x}\right\rangle $
in Eq.~(\ref{eq:U_poly}) with Wick's theorem. As an example, we
can see that the second order of principal minors of $\det\tilde{G}_{x}$
are $\alpha_{i}\alpha_{j}\left|\begin{array}{cc}
\left\langle \hat{n}_{i}\right\rangle  & \left\langle c_{i}^{\dagger}c_{j}\right\rangle \\
\left\langle c_{j}^{\dagger}c_{i}\right\rangle  & \left\langle \hat{n}_{j}\right\rangle 
\end{array}\right|$, with $1\le i<j\le N$, which coincide with $\sum_{1\le i<j\le N}\alpha_{i}\alpha_{j}\left\langle \hat{n}_{i}\hat{n}_{j}\right\rangle $
in Eq.~(\ref{eq:U_poly}) by using Wick's theorem to the latter form.
Furthermore, assuming eigenvalues for $\tilde{G}_{x}$ are $\lambda_{i}$
($i=1,2,\cdots,N$), we have an identity for $\left\langle \hat{U}_{x}\right\rangle $
with eigenvalues as~\citep{meyer_matrix_2023}, 
\begin{equation}
\left\langle \hat{U}_{x}\right\rangle =1+\sum_{l=1}^{N}\lambda_{i}+\sum_{1\le i<j\le N}\lambda_{i}\lambda_{j}+\cdots.
\end{equation}
Compared with the characteristic polynomials $\det\left(\tilde{G}_{x}-\lambda I\right)=\left(-\lambda\right)^{N}+\left(-\lambda\right)^{N-1}\sum_{i=1}^{N}\lambda_{i}+\left(-\lambda\right)^{N-2}\sum_{1\le i<j\le N}\lambda_{i}\lambda_{j}+\cdots$,
and setting $\lambda=-1$, we find 
\begin{equation}
\left\langle \hat{U}_{x}\right\rangle =\det\left(\tilde{G}_{x}+I\right).
\end{equation}
Therefore, finally we have the polarization in the formalism of Green's
functions, 
\begin{equation}
P_{x}=\frac{1}{2\pi\mathcal{S}}\mathrm{Im}\left[\mathrm{Tr}\ln\left(\tilde{G}_{x}+I\right)\right]-P_{bg}.\label{eq:lndetPx}
\end{equation}
With the identity $\mathrm{Tr}\ln A=\ln\det A$, we have obtained
Eq.~(\ref{eq:Px}). We have a remark here for the numerical implementation.
As is known that polarization is an intensive quantity. In practical
calculations, one cannot adopt Eq.~(\ref{eq:lndetPx}), of which
the first term tends to go to zero in the thermodynamic limit in the
more than one-dimensional space~\citep{resta_dipole_2021}, because
numerically the imaginary part of the logarithm function lies in the
interval $\left[-\pi,\pi\right)$. 

Similarly, we can write down the formula for the quadrupole moment
in a crystal, 
\begin{equation}
Q_{xy}=\frac{1}{2\pi\mathcal{L}}\mathrm{Im}\left[\mathrm{Tr}\ln\left(\tilde{G}_{xy}+I\right)\right]-Q_{bg}.\label{eq:Qxy}
\end{equation}
Here $\tilde{G}_{xy}=\alpha_{xy}G$, and $\alpha_{xy}=\mathrm{diag}\{e^{\frac{2\pi ix_{1}y_{1}}{L_{x}L_{y}}}-1,\cdots,e^{\frac{2\pi ix_{N}y_{N}}{L_{x}L_{y}}}-1\}$.
$\mathcal{L}=1$, $L_{z}$ for two- and three-dimensional case, which
is the length perpendicular to the $x-y$ plane. $Q_{bg}=\frac{n_{f}}{\mathcal{L}}\sum_{i}\frac{x_{i}y_{i}}{L_{x}L_{y}}$.
A generalization to the case of the octupole moment is straightforward. 

\begin{figure*}[t]
\includegraphics[scale=0.45]{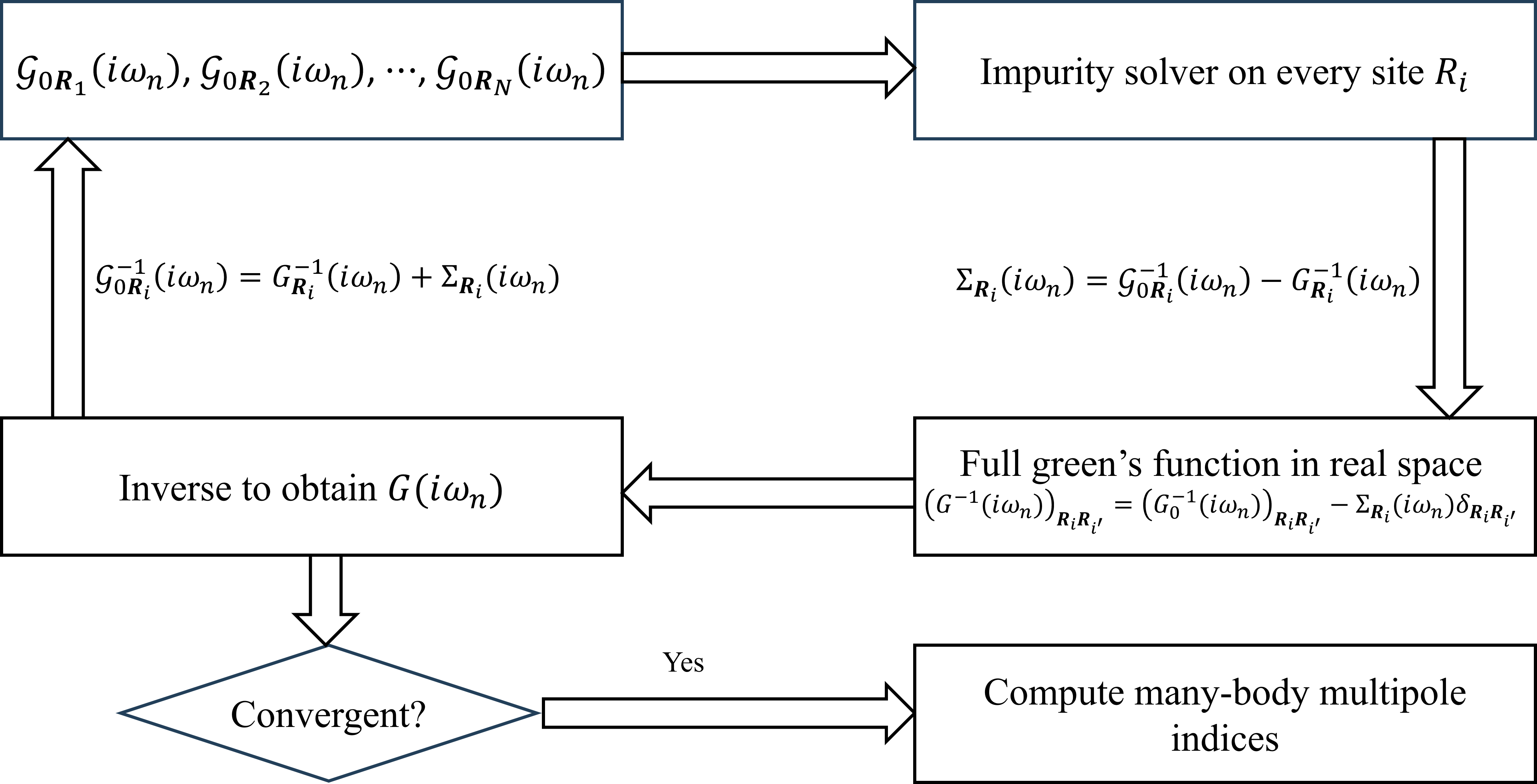}

\caption{\label{fig:Flowchart}Flowchart to compute the many-body multipole
indices with the real-space dynamical mean-field theory (R-DMFT) and
the Green's function formula for the multipole indices in the real-space.
The R-DMFT is featured in Anderson impurities on every lattice site,
self-consistently dealt with impurity solver to obtain $\Sigma_{\bm{R}_{i}}\left(i\omega_{n}\right)$
on site $\bm{R}_{i}$, and meanwhile, the hopping taken into account
by $\left(G_{0}^{-1}\left(i\omega_{n}\right)\right)_{\bm{R}_{i}\bm{R}_{i^{\prime}}}$.
The computation of the multipole indices takes place when the R-DMFT
loop is convergent with the Weiss field $\mathcal{G}_{0\bm{R}_{i}}\left(i\omega_{n}\right)$. }

\end{figure*}

The key issue in the calculations in Eqs.~(\ref{eq:Px}) and~(\ref{eq:Qxy})
is to obtain the matrix of Green's function $G$, and we are interested
in correlated systems described by the fermionic Hubbard model. It
is necessary to obtain $G$ of a correlated lattice model for calculating
of many-body multipole moments. Real-space dynamical mean-field theory
(R-DMFT)~\citep{georges_dynamical_1996,potthoff_self-energy-functional_2003,helmes_dynamical_2008,hofstetter_quantum_2018}
is perfectly suitable for this purpose. In this scheme, a strongly
correlated system on a lattice is mapped into a group of impurities
on lattice sites with the same spatial symmetry, along with hopping
between different sites. In this way both the onsite interactions
and the inhomogeneous due to hopping are taken into account. One then
solves the DMFT problem on all the sites until it is fully converged.
Therefore, the interaction effects are taken into account on the DMFT
level. Once the R-DMFT loop is converged, we then adopt Eqs.~(\ref{eq:Px})
and~(\ref{eq:Qxy}) to compute multipole indices. The flow chart
for the whole computation process is shown in Fig.~\ref{fig:Flowchart},
which can be mostly divided into two parts: (1) the R-DMFT part, and
(2) the computation of many-body multipole indices. We comment on
the efficiency of this algorithm. The most time-consuming part would
be the R-DMFT routines, which can be accelerated with parallel implementation
and optimized by the spatial symmetry consideration. With convergent
Green's function, the calculations of multipole indices are quite
efficient. 

To sum up, we have achieved two significant advancements. (1) We have
a formula in the formalism of real-space Green's functions for the
multipole indices, which is ready to be combined with the DMFT method,
and one can further introduce disorders to lattices to explore the
rich physics of disorder effects, quasi-crystal lattice~\citep{tran_topological_2015},
amorphous lattices~\citep{agarwala_topological_2017,yang_topological_2019}
and so on. (2) With the multipole indices as a bridge, one can certainly
study the localization length of the electron~\citep{resta_electron_1999},
the polarization fluctuation~\citep{resta_polarization_2006}, and
the quantum geometric tensor~\citep{souza_polarization_2000} in
the correlated lattices. 

\section{\label{sec:R-DMFT-calculation}Demonstrating R-DMFT calculations
of the interacting BBH model with staggered potentials }

In this section, we implement the multipole moment of the correlated
BBH model based on Eqs.~(\ref{eq:Px}) and~(\ref{eq:Qxy}) with
the real-space Green's function calculated by the R-DMFT, and discuss
the correlation effects explicitly. 

\subsection{The interacting BBH Model}

The interacting BBH model~\citep{benalcazar_quantized_2017,benalcazar_electric_2017}
on a two-dimensional lattice with the Hubbard interaction and staggered
potentials reads, 

\begin{align*}
H & =H_{0}+H_{stagger}+U\sum_{\bm{R}\gamma}n_{\bm{R}\gamma\uparrow}n_{\bm{R}\gamma\downarrow},
\end{align*}
where 
\begin{align*}
H_{0} & =\sum_{\bm{R}\sigma}\left[\gamma_{x}\left(c_{\bm{R}1\sigma}^{\dagger}c_{\bm{R}3\sigma}+c_{\bm{R}2\sigma}^{\dagger}c_{\bm{R}4\sigma}\right)\right.\\
 & +\gamma_{y}\left(c_{\bm{R}1\sigma}^{\dagger}c_{\bm{R}4\sigma}-c_{\bm{R}2\sigma}^{\dagger}c_{\bm{R}3\sigma}\right)\\
 & +\lambda_{x}\left(c_{\bm{R}1\sigma}^{\dagger}c_{\bm{R}+\hat{x}3\sigma}+c_{\bm{R}4\sigma}^{\dagger}c_{\bm{R}+\hat{x}2\sigma}\right)\\
 & \left.+\lambda_{y}\left(c_{\bm{R}1\sigma}^{\dagger}c_{\bm{R}+\hat{y}4\sigma}-c_{\bm{R}3\sigma}^{\dagger}c_{\bm{R}+\hat{y}2\sigma}\right)\right]+H.c.,
\end{align*}
and $H_{stagger}$ is the onsite staggered potential. The non-interacting
part $H_{0}$ is a standard toy model of the higher-order topological
insulator. In general, spatial symmetries are closely related the
quantization of the polarization and quadrupole moments~\citep{benalcazar_quantized_2017,benalcazar_electric_2017}.
By the term $H_{stagger}$ we can tune the spatial symmetry, and explore
the correlating effects by introducing the Hubbard interaction term.
We are mainly interested into two cases with different spatial symmetries
as follows: case (i) $H_{stagger}=\Delta_{xy}\sum_{\bm{R}\sigma}\left(n_{\bm{R}1\sigma}+n_{\bm{R}2\sigma}-n_{\bm{R}3\sigma}-n_{\bm{R}4\sigma}\right)$
preserving the spatial inversion symmetry, and case (ii) $H_{stagger}=\Delta_{x}\sum_{\bm{R}\sigma}\left(n_{\bm{R}1\sigma}-n_{\bm{R}2\sigma}-n_{\bm{R}3\sigma}+n_{\bm{R}4\sigma}\right)$
or $H_{stagger}=\Delta_{y}\sum_{\bm{R}\sigma}\left(n_{\bm{R}1\sigma}-n_{\bm{R}2\sigma}+n_{\bm{R}3\sigma}-n_{\bm{R}4\sigma}\right)$
which respects mirror symmetry along the $y$-direction (or the $x$-direction). 

\subsection{Results and discussion }

As a demonstration for our general protocol to calculate the multipole
indices in correlated insulators, we present results for the polarization
and quadrupole moments of the interacting BBH model when the onsite
staggered potential is nonzero, which affords us a way to tune the
spatial symmetry. For the R-DMFT part calculations, an impurity solver
of iterative perturbation theory~\citep{georges_dynamical_1996}
is adopted. We choose the periodical boundary conditions for lattices,
work at the half-filling with $n_{f}=\frac{1}{2}$, and focus on the
paramagnetic phase. As a side remark, we point out that all the multipole
moments data are presented by modulo 1. 

\subsubsection{Case (i)}

\begin{figure*}[h]
\includegraphics[scale=0.32]{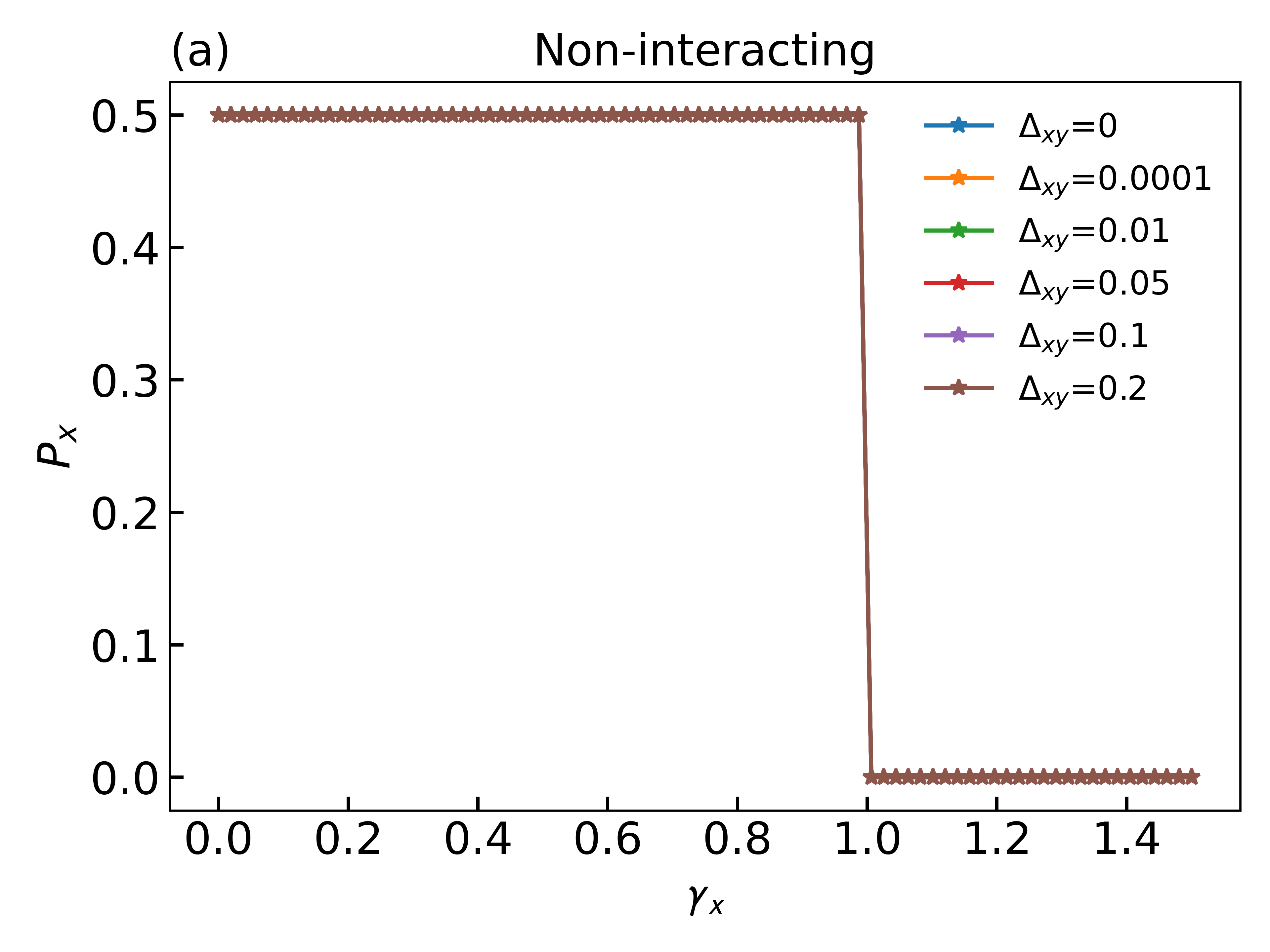}\includegraphics[scale=0.32]{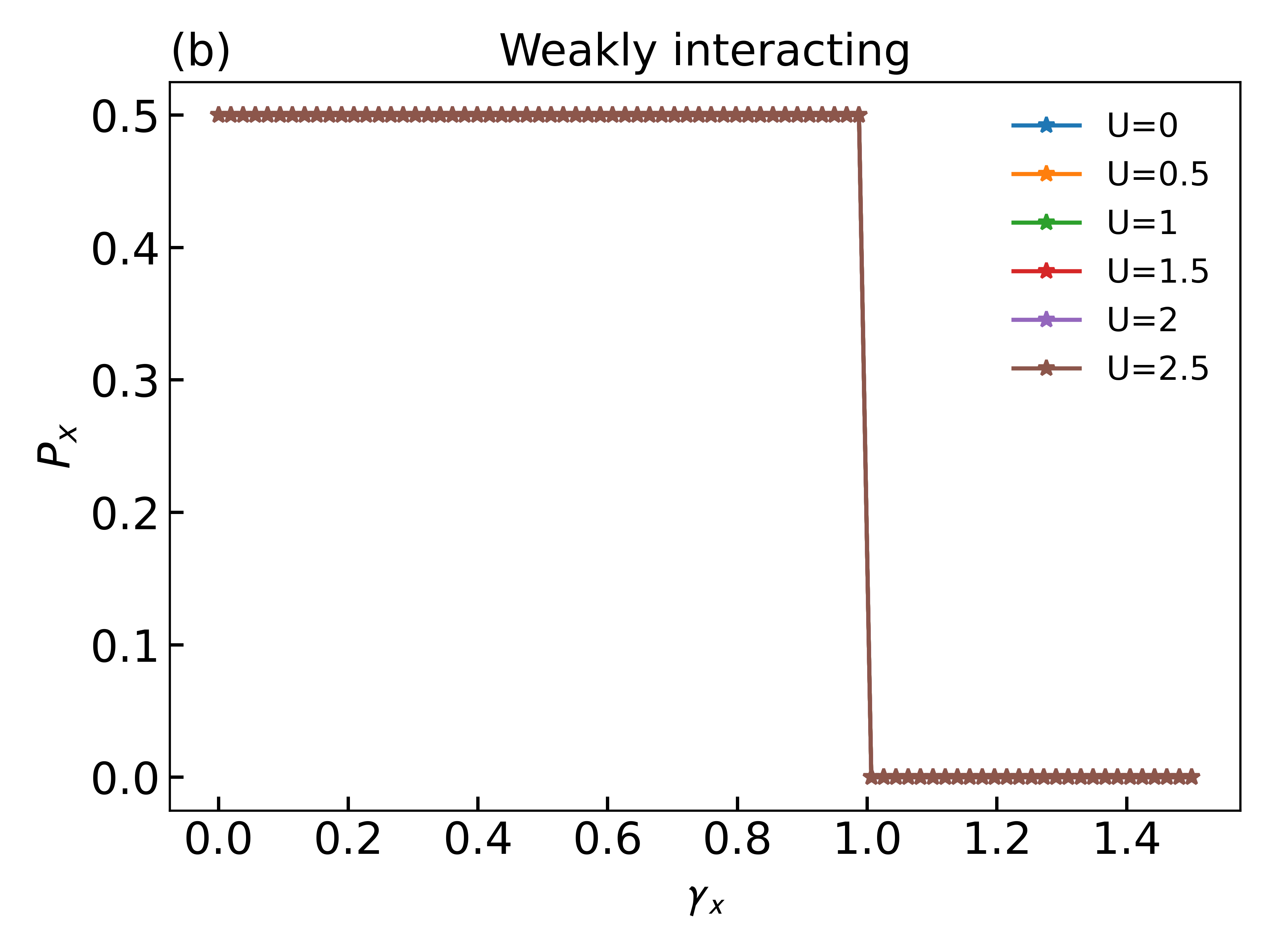}\includegraphics[scale=0.32]{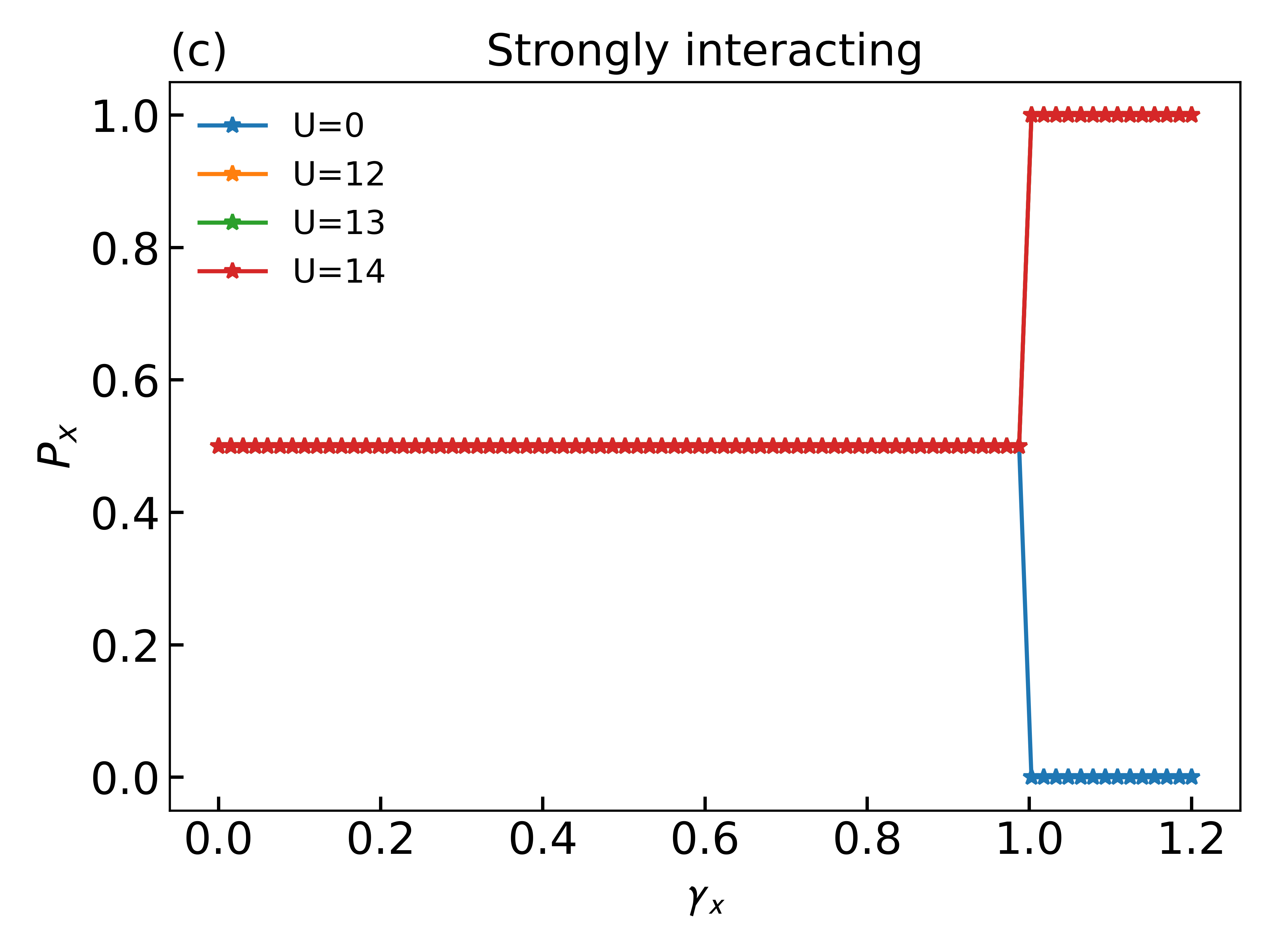}

\includegraphics[scale=0.32]{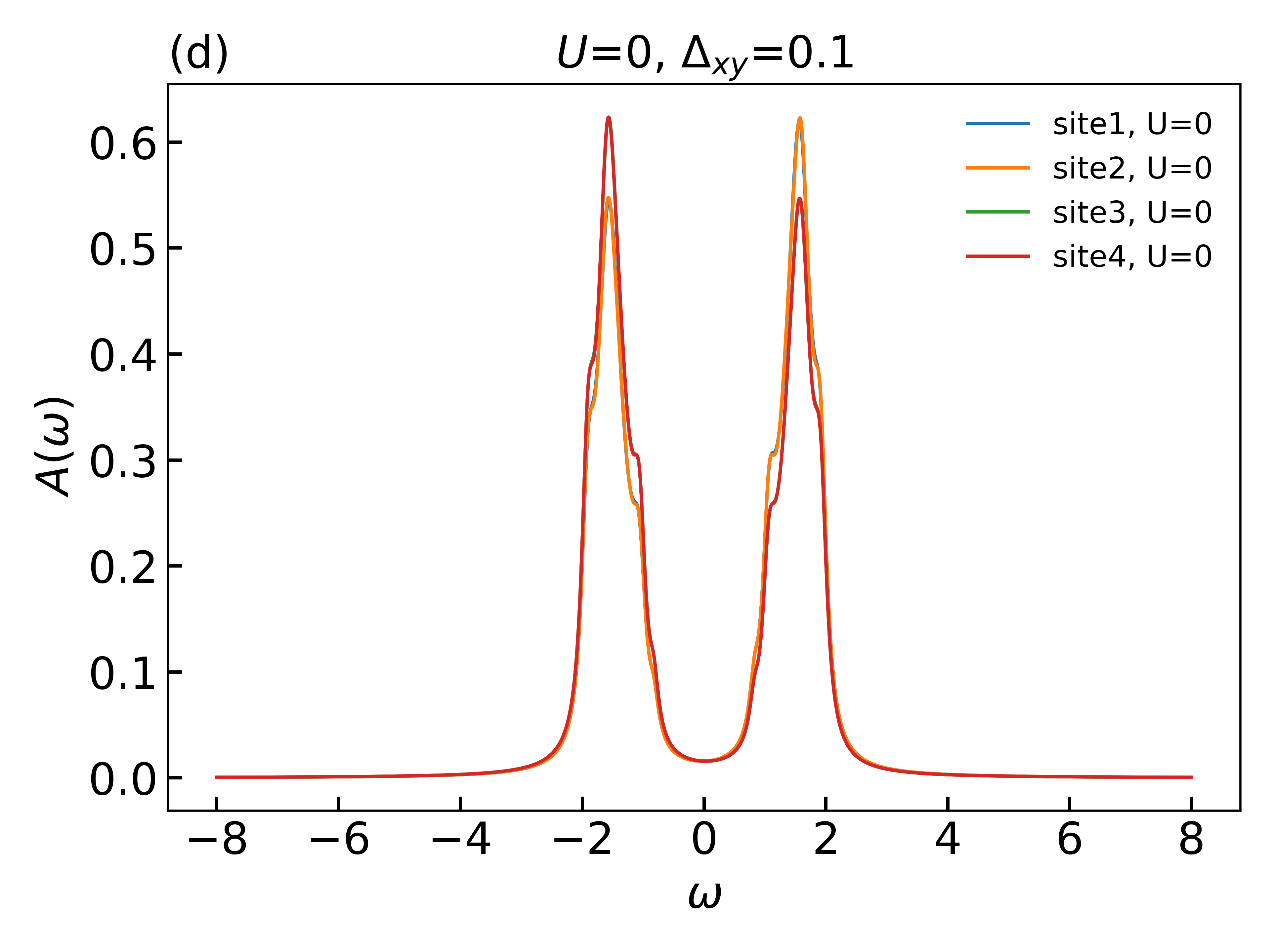}\includegraphics[scale=0.32]{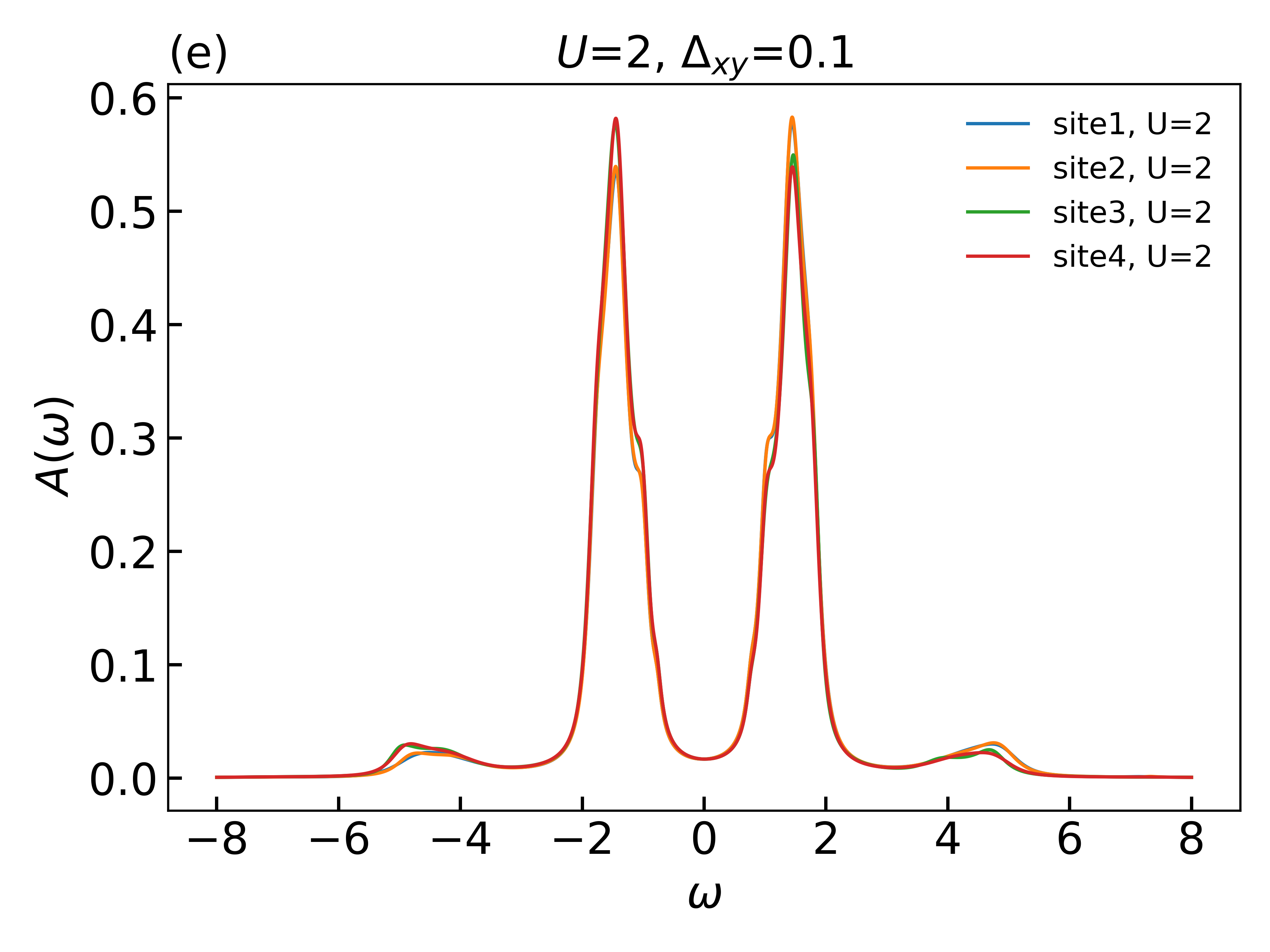}\includegraphics[scale=0.32]{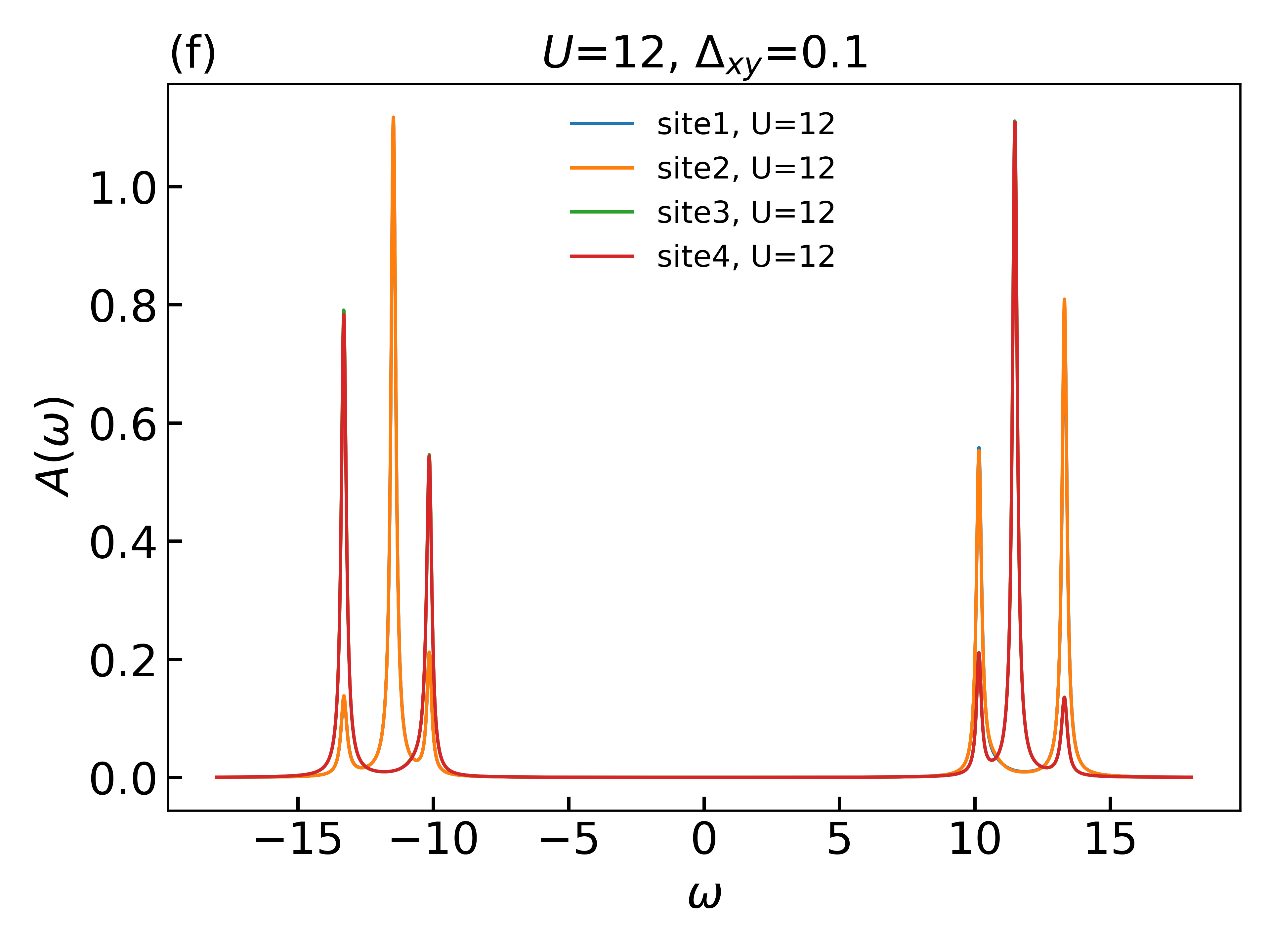}

\caption{\label{fig:casei}Case (i): (a-c) The polarization component $P_{x}$
versus $\gamma_{x}$($\gamma_{y}$) of the BBH model for case (i)
versus the staggered potential $\Delta_{xy}$ and the interaction
strength $U$, with a lattice size of $12\times12$ and $\gamma_{y}=\gamma_{x}$,
and (d-f) spectral functions for different interactions with $\gamma_{x}=\gamma_{y}=0.4$.
We have $\lambda_{x}=\lambda_{y}=1$ as the energy unit. In panel
(a) we explore different $\Delta_{xy}$ when $U=0$. And with a fixed
$\Delta_{xy}=0.1$ we present $P_{x}$ for weak interaction in panel
(b) and strong interaction in panel (c). For the DMFT calculations,
$\beta=50$, the number of Matsubara frequencies is $256$, and the
Pade approximation~\citep{vidberg_solving_1977} is adopted in the
analytical continuation of Green's function to the real frequency.}
\end{figure*}

As shown in Fig.~\ref{fig:casei}, there are quantized polarization
component $P_{x}$ and sharp topological phase transitions in non-interacting
case, which are features of the BBH model, and remarkably similar
phenomena in interacting regimes. The results given by the formula~(\ref{eq:Px})
are consistent with the spatial symmetry analysis. In Fig.~\ref{fig:casei},
we do not present results for $P_{y}$, whose features are quite similar
to these for $P_{x}$. The onsite staggered potential $\Delta_{xy}$
breaks the reflection symmetries about both $x$ and $y$ axes, but
keeps the inversion symmetry, which leaves the polarization quantized
in Fig.~\ref{fig:casei} for non-interacting case (a), weakly interacting
case (b) and strongly interacting case (c). Within the R-DMFT method,
even though the interaction is turned on, the polarization component
$P_{x}$ keep quantized as long as the spatial symmetry conditions
for quantization are preserved. In Fig.~\ref{fig:casei}, we show
the spectral functions for four different sites in one unit cell,
and the gaps due to interactions together with emergent side-bands
in Figs.~\ref{fig:casei}(e-f) for very strong interactions clearly
show that the system is driven into the Mott insulator phase when
the interaction $U$ is large enough. To sum up, we show that even
in the correlating regime, the quantization of polarization is still
determined by the spatial symmetries. 

\begin{figure*}[h]
\includegraphics[scale=0.32]{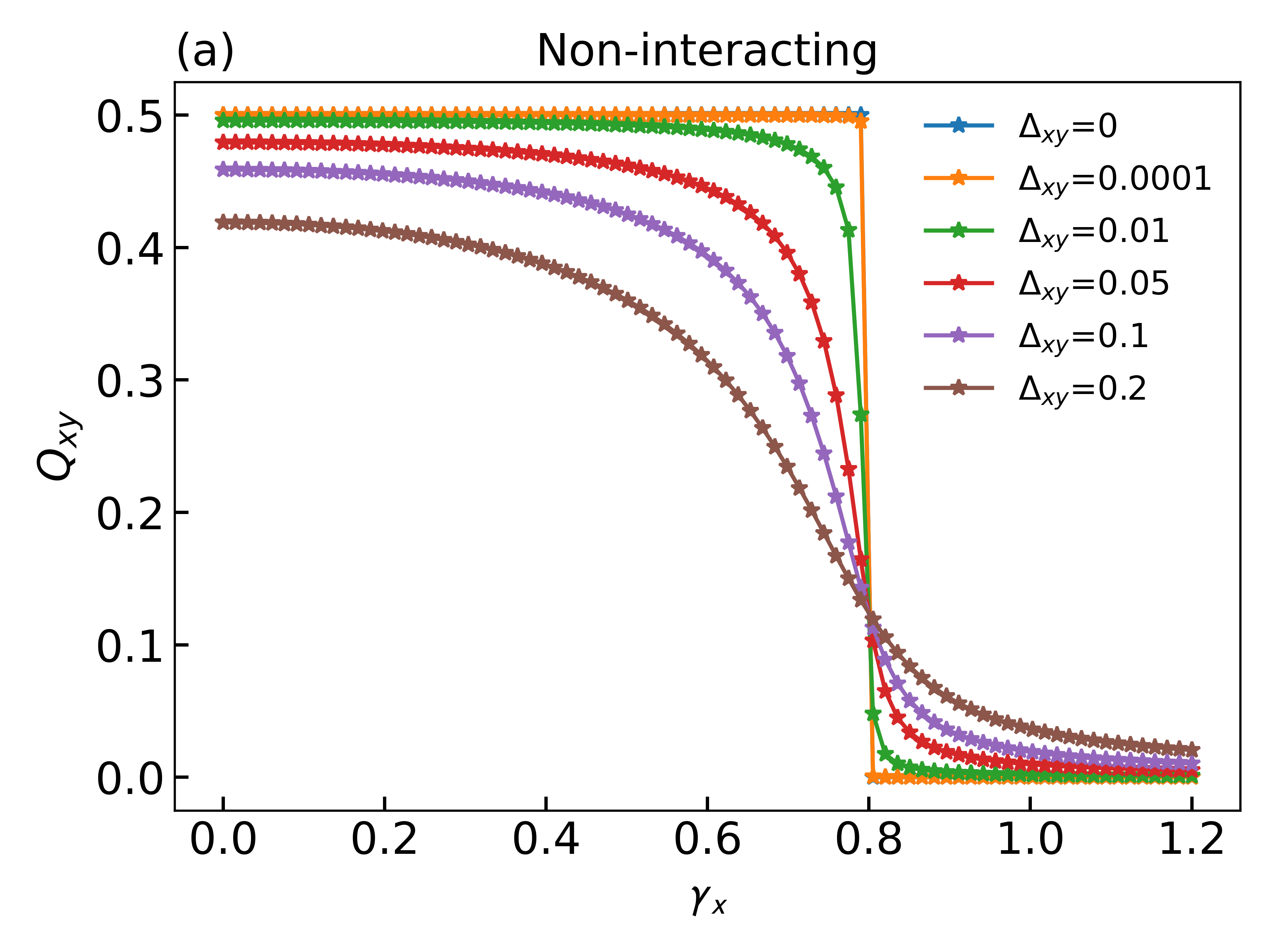}\includegraphics[scale=0.32]{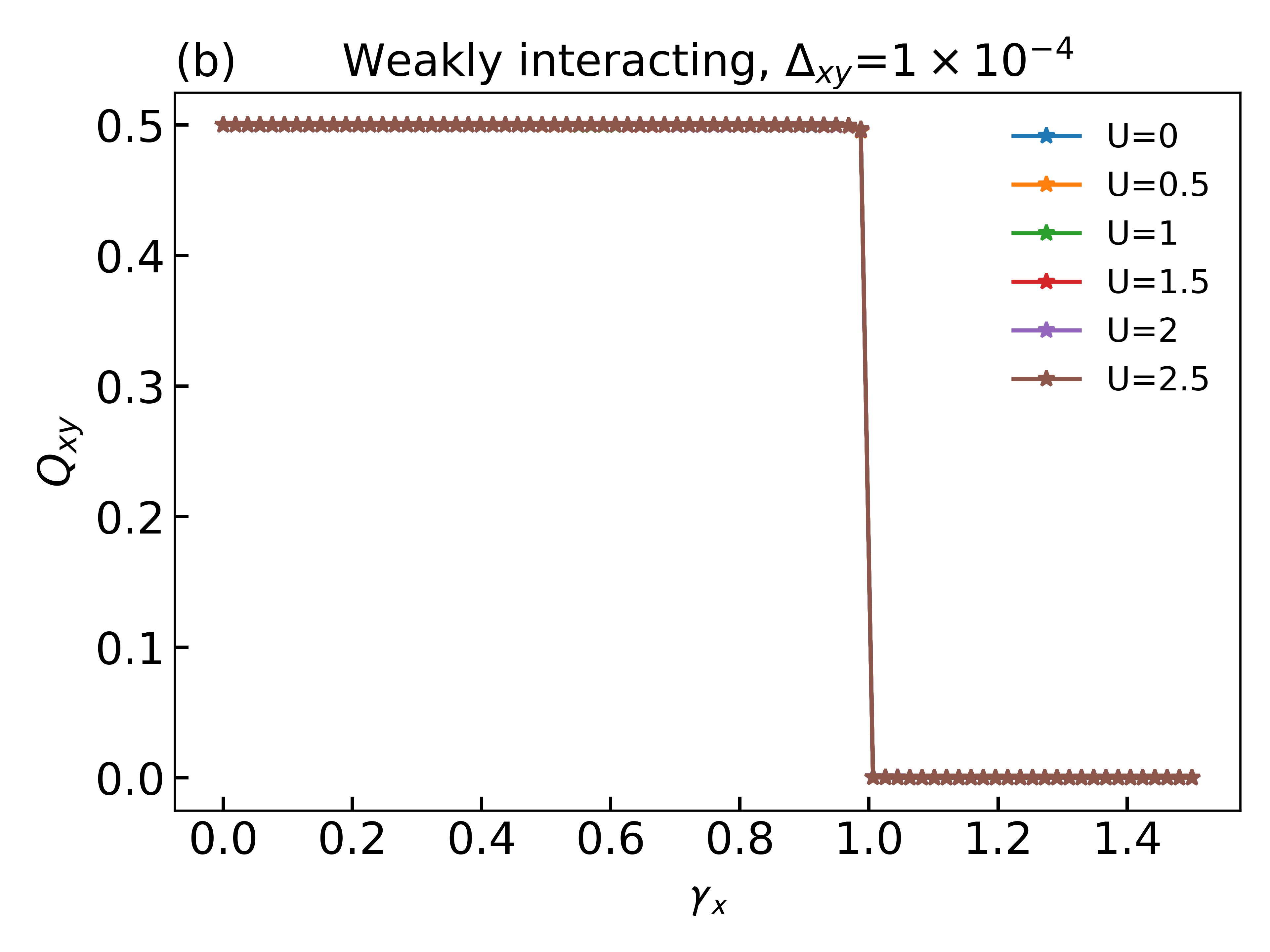}\includegraphics[scale=0.32]{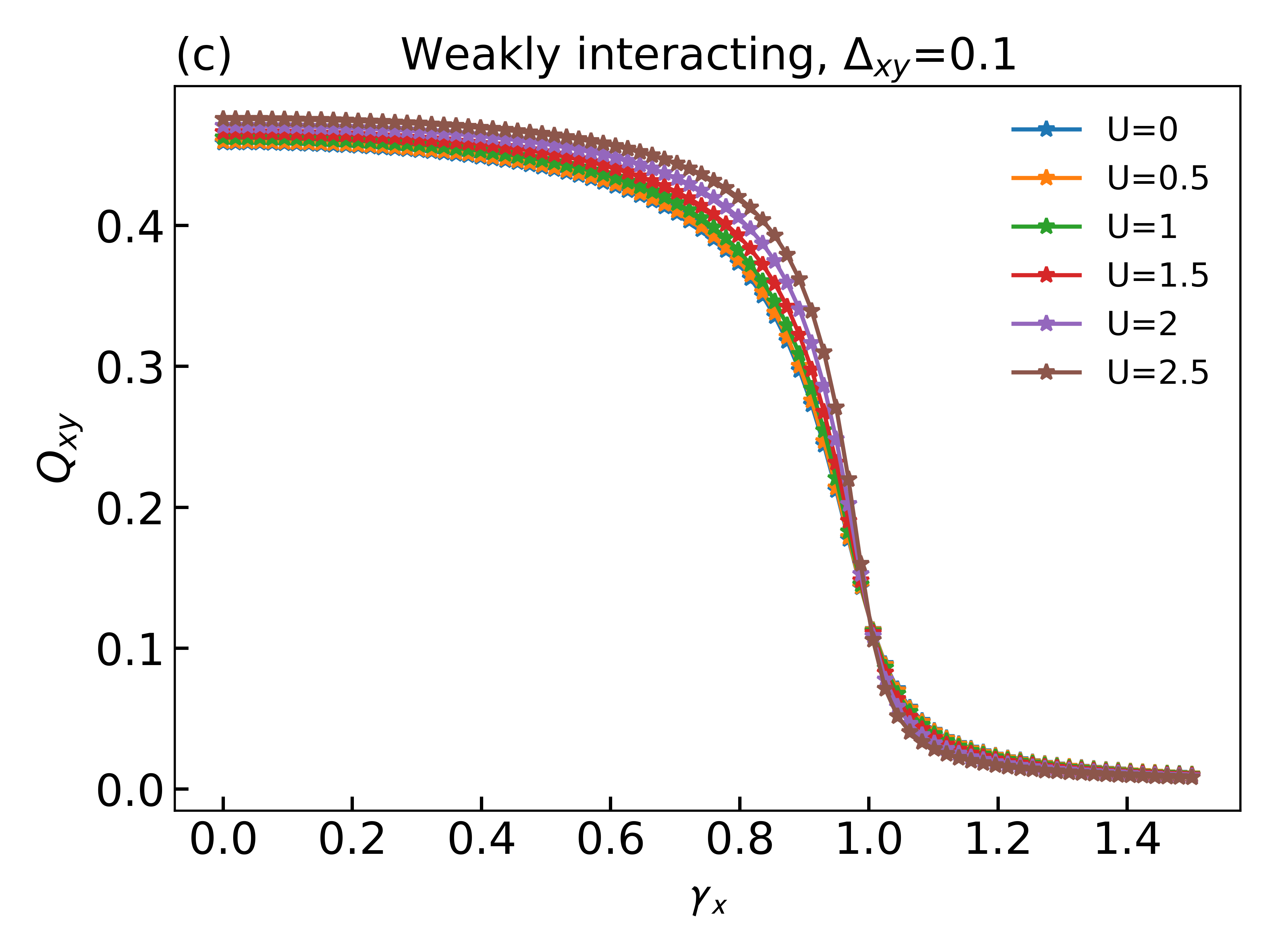}

\caption{\label{fig:quadrupole}Case (i): The quadrupole moment $Q_{xy}$ versus
$\gamma_{x}$($\gamma_{y}$) of the BBH model with a lattice size
of $12\times12$ for (a) noninteracting case with different strengths
of $\Delta_{xy}$, (b) interacting case with $\Delta_{xy}=1\times10^{-4}$,
and (c) interacting case with $\Delta_{xy}=0.1$. $\gamma_{x}=\gamma_{y}$.
Other parameters are the same as these in Fig.~\ref{fig:casei}.}
\end{figure*}

Furthermore, we present our results for the quadrupole moment calculations
with Eq.~(\ref{eq:Qxy}) in Fig.~\ref{fig:quadrupole}. If the reflection
symmetries are absent, the quadrupole moment $Q_{xy}$ is unquantized
in Figs.~\ref{fig:quadrupole}(a) with nonzero $\Delta_{xy}$, and
(c) with both $\Delta_{xy}$ and interaction $U$ turned on. As long
as the reflection symmetry is preserved in both $x$ and $y$ directions,
the quadrupole moment stays quantized as shown in Fig.~\ref{fig:quadrupole}(b).
It shows that the spatial symmetry is a dominant factor for the quantization
of the quadrupole moment in the paramagnetic phase. 

\subsubsection{Case (ii)}

\begin{figure*}[h]
\includegraphics[scale=0.32]{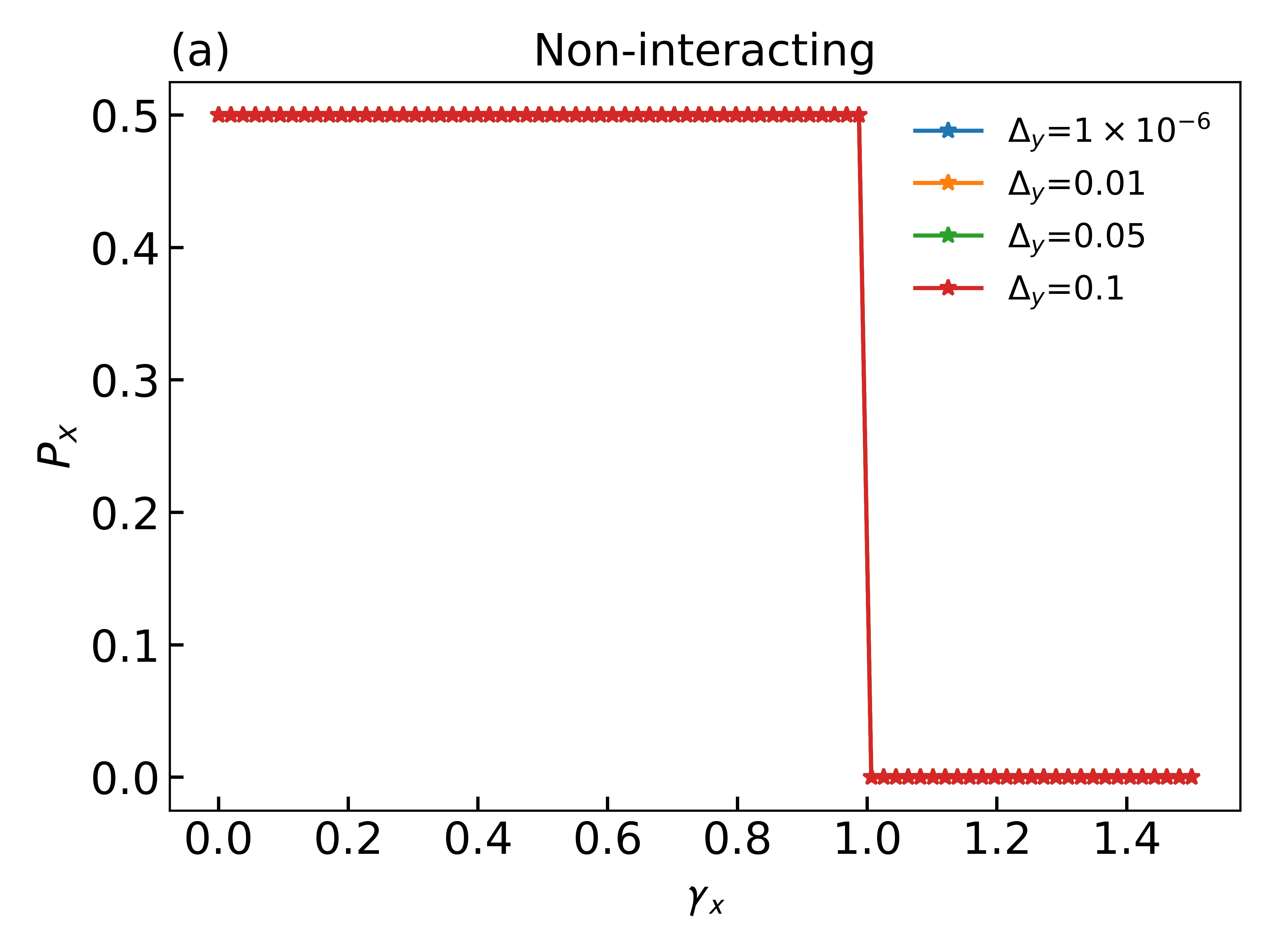}\includegraphics[scale=0.32]{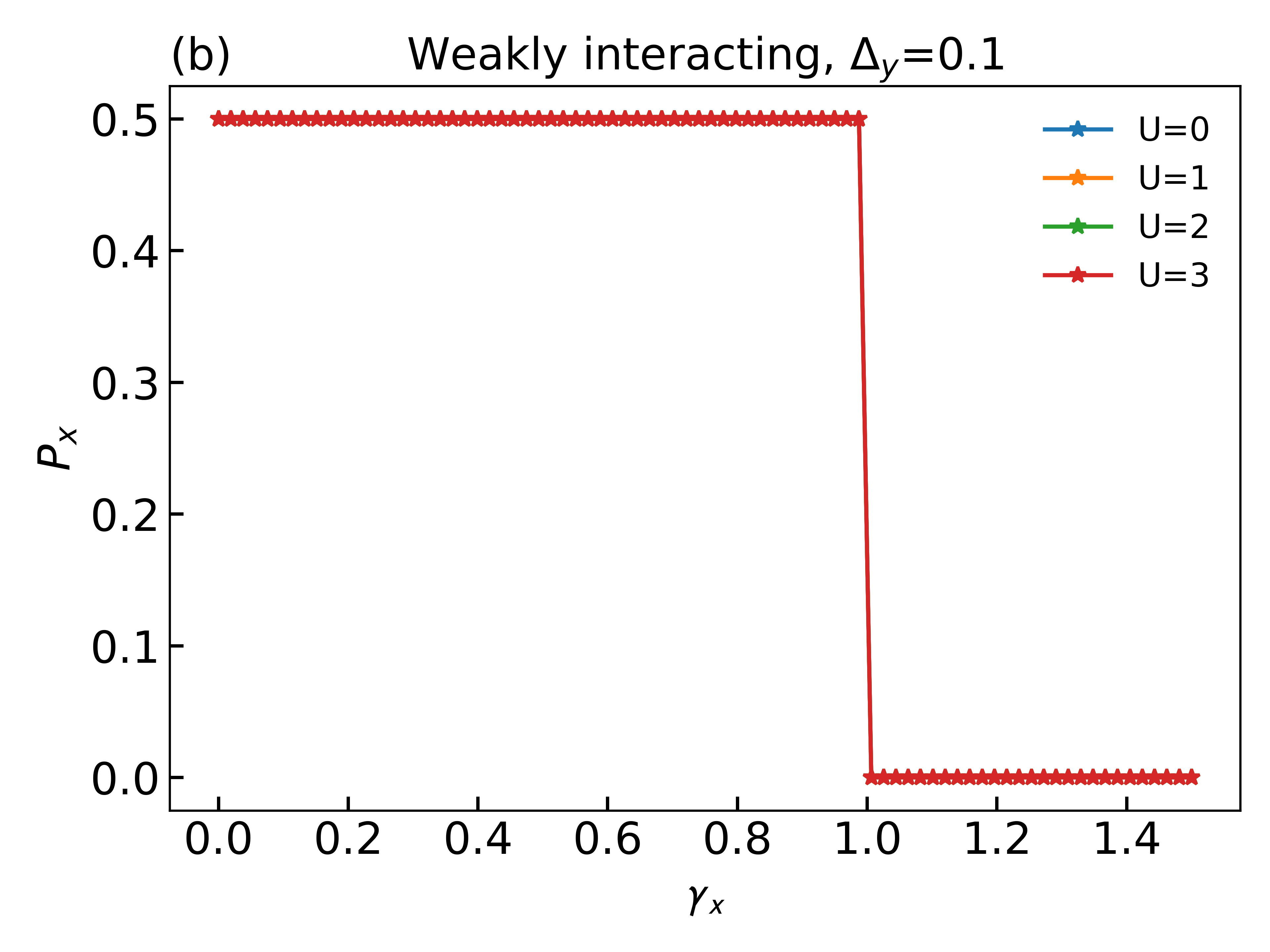}\includegraphics[scale=0.32]{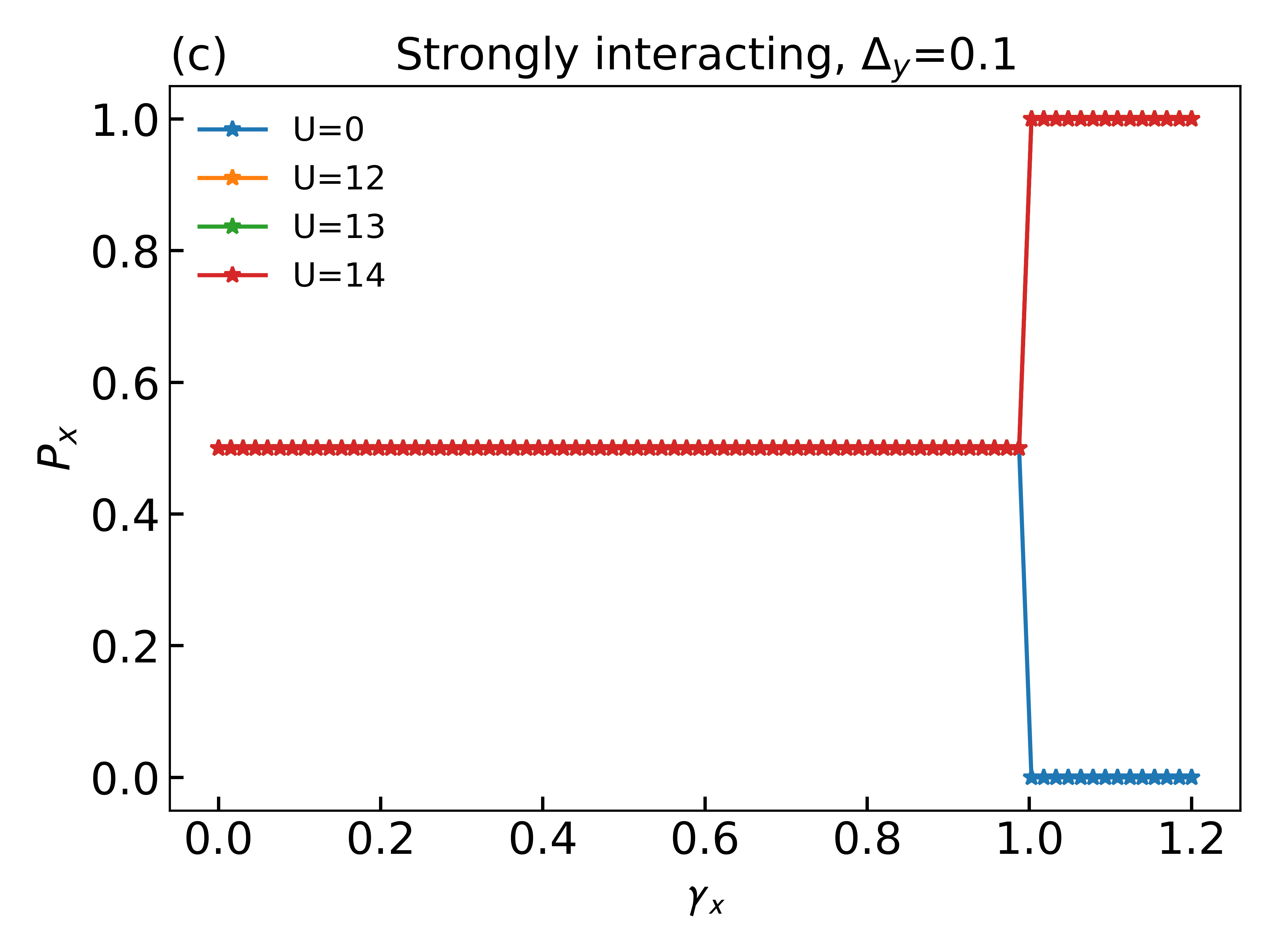}

\includegraphics[scale=0.32]{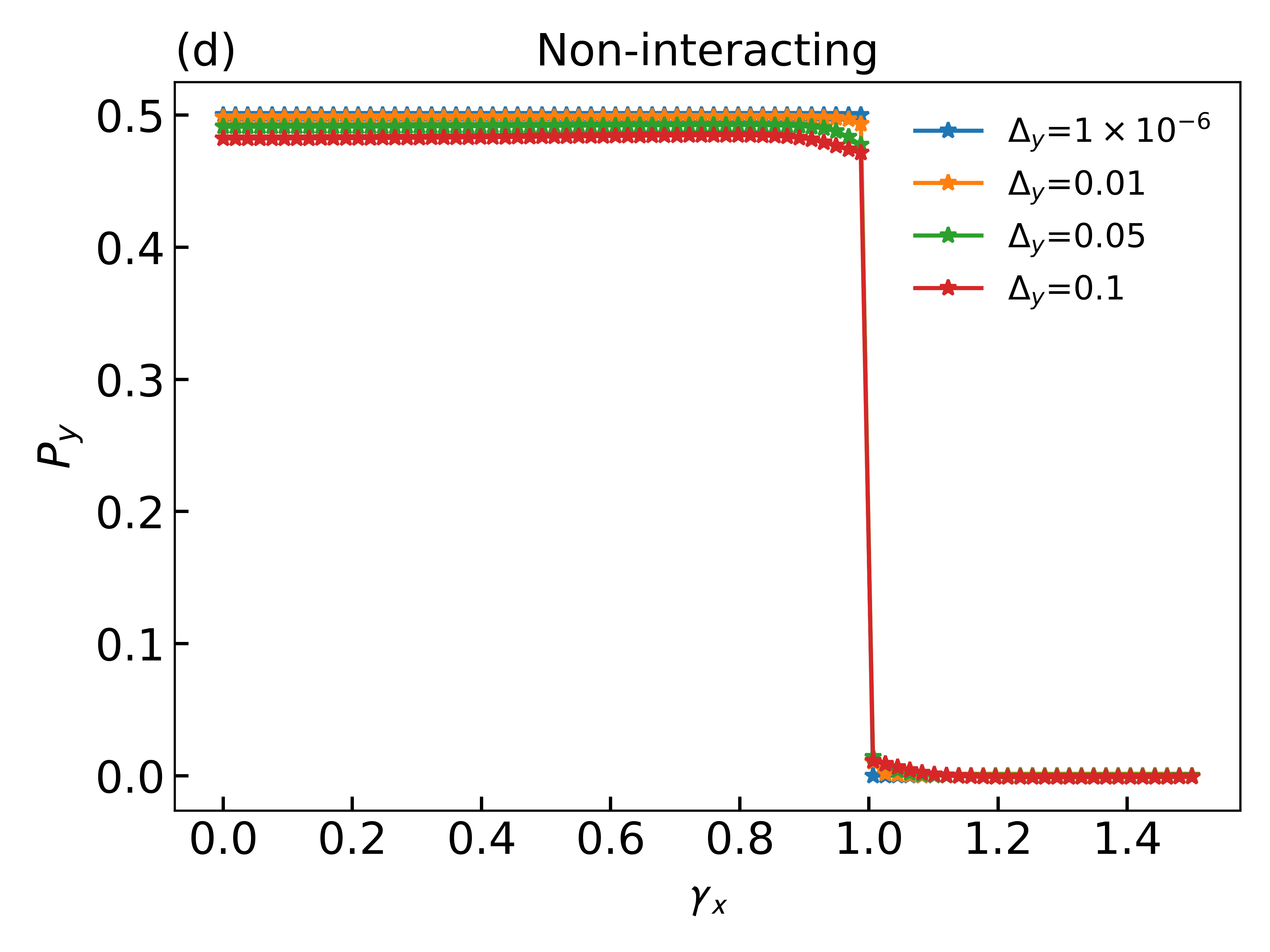}\includegraphics[scale=0.32]{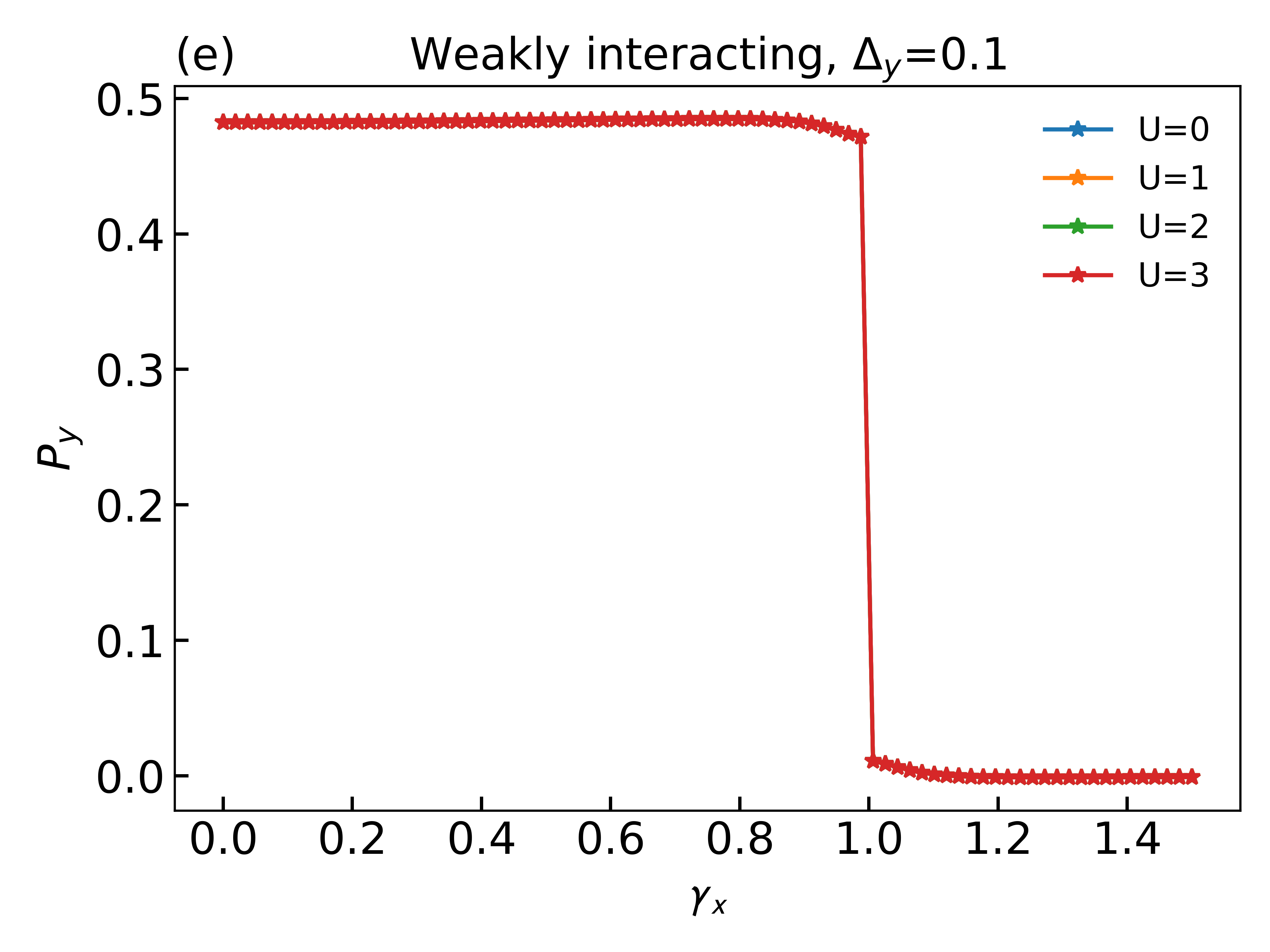}\includegraphics[scale=0.32]{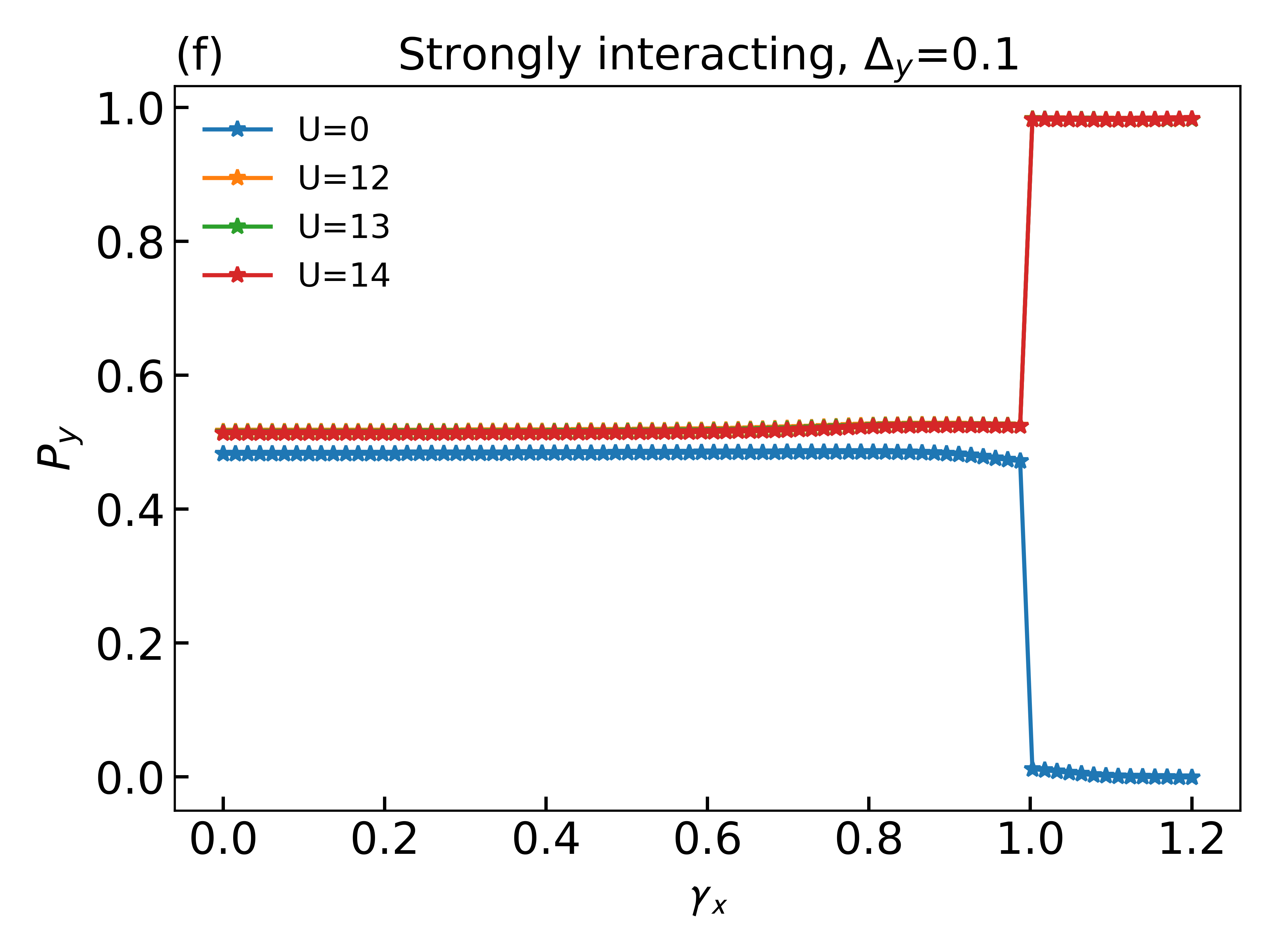}

\caption{\label{fig:caseii}Case (ii): The polarization components $P_{x}$
and $P_{y}$ versus $\gamma_{x}$($\gamma_{y}$) of the BBH model
with a lattice size of $12\times12$. We present results for $P_{x}$
and $P_{y}$, in (a, d) for non-interacting case with different strengths
of $\Delta_{y}$, (b, e) for weak interactions with finite $\Delta_{y}$
and (c, f) for strong interaction with finite $\Delta_{y}$. $\gamma_{x}=\gamma_{y}$.
Other parameters are the same as these in Fig.~\ref{fig:casei}. }
\end{figure*}

In this subsection, we present our results for the polarization $P_{x}$
and $P_{y}$ for case (ii) with $H_{stagger}=\Delta_{y}\sum_{\bm{R}\sigma}\left(n_{\bm{R}1\sigma}-n_{\bm{R}2\sigma}+n_{\bm{R}3\sigma}-n_{\bm{R}4\sigma}\right)$,
which respects the mirror symmetry in the $x$ direction. In Fig.~\ref{fig:caseii}(a)
for non-interacting case with different strengths of $\Delta_{y}$,
(b) for weak interactions with finite $\Delta_{y}$ and (c) for strong
interaction with finite $\Delta_{y}$, we observe perfect quantization
of the polarization $P_{x}$ and clear topological transition when
both the interaction and $\Delta_{y}$ is finite. Due to the spatial
symmetry of case (ii), the topological transition here is in fact
quite similar to that in the one-dimensional Su-Schrieffer-Heeger
model~\citep{su_soliton_1980}. As a sharp contrast in Figs.~\ref{fig:caseii}(d-f),
we can see the quantization of $P_{y}$ is broken once $\Delta_{y}$
is nonzero, indicating the broken of the mirror symmetry. By the comparison
between $P_{x}$ and $P_{y}$, we clearly see that the spatial symmetry
is a vital factor for the quantization of the polarization in the
paramagnetic phase. As a side comment, in Figs.~\ref{fig:caseii}(a,
d), we introduce a tiny $\Delta_{x}=1\times10^{-6}$ to break the
mirror symmetry with the $x$ axis a little, as a numerical trick
to remove the ``degeneracy'' between $-0.5$ and $0.5$ for the
polarization in the sense of modulo~\citep{benalcazar_electric_2017}. 

We have a few words on the experimental measurement. The prediction
of the multipole moments is ready to be checked by experiment measurements.
Even though it is difficult to realize the BBH model in real materials,
it seems promising to do it in metamaterials~\citep{yatsugi_observation_2022,dong_topolectric_2021-1}.

\section{\label{sec:Outlook}Summary and outlook }

To sum up, we propose a practical calculation scheme for the multipole
moments in correlated materials on the level of R-DMFT. By calculations
of BBH with different spatial symmetries due to staggered potentials,
and different strengths of interactions we show the effectiveness
of our method in exploring the properties of the multipole indices.
However, its full potential is not fully unleashed. Our general method
is ready to be implemented to explore a disordered correlated lattice.
The close relation of polarization with the localization length~\citep{resta_electron_1999},
and the quantum geometric tensor~\citep{souza_polarization_2000}
affords us a way to explore these interesting physical quantities.
Our demonstrating calculations show that the spatial symmetry is a
dominant factor for the quantization of the polarization and quadrupole
moments even in the correlating regime, at least in the paramagnetic
phase. 
\begin{acknowledgments}
T.Q. and J.H.Z. were supported by the National Natural Science Foundation
of China under Grants No.12174394 and U2032164. J.H.Z. was also supported
by HFIPS Director\textquoteright s Fund (Grants No. YZJJQY202304 and
No. BJPY2023B05) and Anhui Provincial Major S\&T Project (s202305a12020005),
the High Magnetic Field Laboratory of Anhui Province under Contract
No. AHHM-FX-2020-02. A portion of this work was supported by Chinese
Academy of Sciences under contract No. JZHKYPT-2021-08.
\end{acknowledgments}

\bibliographystyle{apsrev4-1}

\begin{thebibliography}{37}%
\makeatletter
\providecommand \@ifxundefined [1]{%
 \@ifx{#1\undefined}
}%
\providecommand \@ifnum [1]{%
 \ifnum #1\expandafter \@firstoftwo
 \else \expandafter \@secondoftwo
 \fi
}%
\providecommand \@ifx [1]{%
 \ifx #1\expandafter \@firstoftwo
 \else \expandafter \@secondoftwo
 \fi
}%
\providecommand \natexlab [1]{#1}%
\providecommand \enquote  [1]{``#1''}%
\providecommand \bibnamefont  [1]{#1}%
\providecommand \bibfnamefont [1]{#1}%
\providecommand \citenamefont [1]{#1}%
\providecommand \href@noop [0]{\@secondoftwo}%
\providecommand \href [0]{\begingroup \@sanitize@url \@href}%
\providecommand \@href[1]{\@@startlink{#1}\@@href}%
\providecommand \@@href[1]{\endgroup#1\@@endlink}%
\providecommand \@sanitize@url [0]{\catcode `\\12\catcode `\$12\catcode
  `\&12\catcode `\#12\catcode `\^12\catcode `\_12\catcode `\%12\relax}%
\providecommand \@@startlink[1]{}%
\providecommand \@@endlink[0]{}%
\providecommand \url  [0]{\begingroup\@sanitize@url \@url }%
\providecommand \@url [1]{\endgroup\@href {#1}{\urlprefix }}%
\providecommand \urlprefix  [0]{URL }%
\providecommand \Eprint [0]{\href }%
\providecommand \doibase [0]{http://dx.doi.org/}%
\providecommand \selectlanguage [0]{\@gobble}%
\providecommand \bibinfo  [0]{\@secondoftwo}%
\providecommand \bibfield  [0]{\@secondoftwo}%
\providecommand \translation [1]{[#1]}%
\providecommand \BibitemOpen [0]{}%
\providecommand \bibitemStop [0]{}%
\providecommand \bibitemNoStop [0]{.\EOS\space}%
\providecommand \EOS [0]{\spacefactor3000\relax}%
\providecommand \BibitemShut  [1]{\csname bibitem#1\endcsname}%
\let\auto@bib@innerbib\@empty
\bibitem [{\citenamefont {Resta}(1992)}]{resta_theory_1992}%
  \BibitemOpen
  \bibfield  {author} {\bibinfo {author} {\bibfnamefont {R.}~\bibnamefont
  {Resta}},\ }\href {\doibase 10.1080/00150199208016065} {\bibfield  {journal}
  {\bibinfo  {journal} {Ferroelectrics}\ }\textbf {\bibinfo {volume} {136}},\
  \bibinfo {pages} {51} (\bibinfo {year} {1992})}\BibitemShut {NoStop}%
\bibitem [{\citenamefont {King-Smith}\ and\ \citenamefont
  {Vanderbilt}(1993)}]{king-smith_theory_1993}%
  \BibitemOpen
  \bibfield  {author} {\bibinfo {author} {\bibfnamefont {R.~D.}\ \bibnamefont
  {King-Smith}}\ and\ \bibinfo {author} {\bibfnamefont {D.}~\bibnamefont
  {Vanderbilt}},\ }\href {\doibase 10.1103/PhysRevB.47.1651} {\bibfield
  {journal} {\bibinfo  {journal} {Phys. Rev. B}\ }\textbf {\bibinfo {volume}
  {47}},\ \bibinfo {pages} {1651} (\bibinfo {year} {1993})}\BibitemShut
  {NoStop}%
\bibitem [{\citenamefont {Resta}(1994)}]{resta_macroscopic_1994}%
  \BibitemOpen
  \bibfield  {author} {\bibinfo {author} {\bibfnamefont {R.}~\bibnamefont
  {Resta}},\ }\href {\doibase 10.1103/RevModPhys.66.899} {\bibfield  {journal}
  {\bibinfo  {journal} {Rev. Mod. Phys.}\ }\textbf {\bibinfo {volume} {66}},\
  \bibinfo {pages} {899} (\bibinfo {year} {1994})}\BibitemShut {NoStop}%
\bibitem [{\citenamefont {Ortiz}\ and\ \citenamefont
  {Martin}(1994)}]{ortiz_macroscopic_1994}%
  \BibitemOpen
  \bibfield  {author} {\bibinfo {author} {\bibfnamefont {G.}~\bibnamefont
  {Ortiz}}\ and\ \bibinfo {author} {\bibfnamefont {R.~M.}\ \bibnamefont
  {Martin}},\ }\href {\doibase 10.1103/PhysRevB.49.14202} {\bibfield  {journal}
  {\bibinfo  {journal} {Phys. Rev. B}\ }\textbf {\bibinfo {volume} {49}},\
  \bibinfo {pages} {14202} (\bibinfo {year} {1994})}\BibitemShut {NoStop}%
\bibitem [{\citenamefont {Coh}\ and\ \citenamefont
  {Vanderbilt}(2009)}]{coh_electric_2009}%
  \BibitemOpen
  \bibfield  {author} {\bibinfo {author} {\bibfnamefont {S.}~\bibnamefont
  {Coh}}\ and\ \bibinfo {author} {\bibfnamefont {D.}~\bibnamefont
  {Vanderbilt}},\ }\href {\doibase 10.1103/PhysRevLett.102.107603} {\bibfield
  {journal} {\bibinfo  {journal} {Phys. Rev. Lett.}\ }\textbf {\bibinfo
  {volume} {102}},\ \bibinfo {pages} {107603} (\bibinfo {year}
  {2009})}\BibitemShut {NoStop}%
\bibitem [{\citenamefont {Xiao}\ \emph {et~al.}(2010)\citenamefont {Xiao},
  \citenamefont {Chang},\ and\ \citenamefont {Niu}}]{xiao_berry_2010}%
  \BibitemOpen
  \bibfield  {author} {\bibinfo {author} {\bibfnamefont {D.}~\bibnamefont
  {Xiao}}, \bibinfo {author} {\bibfnamefont {M.-C.}\ \bibnamefont {Chang}}, \
  and\ \bibinfo {author} {\bibfnamefont {Q.}~\bibnamefont {Niu}},\ }\href
  {\doibase 10.1103/RevModPhys.82.1959} {\bibfield  {journal} {\bibinfo
  {journal} {Rev. Mod. Phys.}\ }\textbf {\bibinfo {volume} {82}},\ \bibinfo
  {pages} {1959} (\bibinfo {year} {2010})}\BibitemShut {NoStop}%
\bibitem [{\citenamefont {Vaidya}\ \emph {et~al.}(2024)\citenamefont {Vaidya},
  \citenamefont {Rechtsman},\ and\ \citenamefont
  {Benalcazar}}]{vaidya_polarization_2024}%
  \BibitemOpen
  \bibfield  {author} {\bibinfo {author} {\bibfnamefont {S.}~\bibnamefont
  {Vaidya}}, \bibinfo {author} {\bibfnamefont {M.~C.}\ \bibnamefont
  {Rechtsman}}, \ and\ \bibinfo {author} {\bibfnamefont {W.~A.}\ \bibnamefont
  {Benalcazar}},\ }\href {\doibase 10.1103/PhysRevLett.132.116602} {\bibfield
  {journal} {\bibinfo  {journal} {Phys. Rev. Lett.}\ }\textbf {\bibinfo
  {volume} {132}},\ \bibinfo {pages} {116602} (\bibinfo {year}
  {2024})}\BibitemShut {NoStop}%
\bibitem [{\citenamefont {Ortiz}\ \emph {et~al.}(1996)\citenamefont {Ortiz},
  \citenamefont {Ordej\'{o}n}, \citenamefont {Martin},\ and\ \citenamefont
  {Chiappe}}]{ortiz_quantum_1996}%
  \BibitemOpen
  \bibfield  {author} {\bibinfo {author} {\bibfnamefont {G.}~\bibnamefont
  {Ortiz}}, \bibinfo {author} {\bibfnamefont {P.}~\bibnamefont {Ordej\'{o}n}},
  \bibinfo {author} {\bibfnamefont {R.~M.}\ \bibnamefont {Martin}}, \ and\
  \bibinfo {author} {\bibfnamefont {G.}~\bibnamefont {Chiappe}},\ }\href
  {\doibase 10.1103/PhysRevB.54.13515} {\bibfield  {journal} {\bibinfo
  {journal} {Phys. Rev. B}\ }\textbf {\bibinfo {volume} {54}},\ \bibinfo
  {pages} {13515} (\bibinfo {year} {1996})}\BibitemShut {NoStop}%
\bibitem [{\citenamefont {Resta}(1998)}]{resta_quantum-mechanical_1998}%
  \BibitemOpen
  \bibfield  {author} {\bibinfo {author} {\bibfnamefont {R.}~\bibnamefont
  {Resta}},\ }\href {\doibase 10.1103/PhysRevLett.80.1800} {\bibfield
  {journal} {\bibinfo  {journal} {Phys. Rev. Lett.}\ }\textbf {\bibinfo
  {volume} {80}},\ \bibinfo {pages} {1800} (\bibinfo {year}
  {1998})}\BibitemShut {NoStop}%
\bibitem [{\citenamefont {Resta}\ and\ \citenamefont
  {Sorella}(1999)}]{resta_electron_1999}%
  \BibitemOpen
  \bibfield  {author} {\bibinfo {author} {\bibfnamefont {R.}~\bibnamefont
  {Resta}}\ and\ \bibinfo {author} {\bibfnamefont {S.}~\bibnamefont
  {Sorella}},\ }\href {\doibase 10.1103/PhysRevLett.82.370} {\bibfield
  {journal} {\bibinfo  {journal} {Phys. Rev. Lett.}\ }\textbf {\bibinfo
  {volume} {82}},\ \bibinfo {pages} {370} (\bibinfo {year} {1999})}\BibitemShut
  {NoStop}%
\bibitem [{\citenamefont {Kohn}(1964)}]{kohn_theory_1964}%
  \BibitemOpen
  \bibfield  {author} {\bibinfo {author} {\bibfnamefont {W.}~\bibnamefont
  {Kohn}},\ }\href {\doibase 10.1103/PhysRev.133.A171} {\bibfield  {journal}
  {\bibinfo  {journal} {Phys. Rev.}\ }\textbf {\bibinfo {volume} {133}},\
  \bibinfo {pages} {A171} (\bibinfo {year} {1964})}\BibitemShut {NoStop}%
\bibitem [{\citenamefont {Benalcazar}\ \emph
  {et~al.}(2017{\natexlab{a}})\citenamefont {Benalcazar}, \citenamefont
  {Bernevig},\ and\ \citenamefont {Hughes}}]{benalcazar_quantized_2017}%
  \BibitemOpen
  \bibfield  {author} {\bibinfo {author} {\bibfnamefont {W.~A.}\ \bibnamefont
  {Benalcazar}}, \bibinfo {author} {\bibfnamefont {B.~A.}\ \bibnamefont
  {Bernevig}}, \ and\ \bibinfo {author} {\bibfnamefont {T.~L.}\ \bibnamefont
  {Hughes}},\ }\href {\doibase 10.1126/science.aah6442} {\bibfield  {journal}
  {\bibinfo  {journal} {Science}\ }\textbf {\bibinfo {volume} {357}},\ \bibinfo
  {pages} {61} (\bibinfo {year} {2017}{\natexlab{a}})}\BibitemShut {NoStop}%
\bibitem [{\citenamefont {Xie}\ \emph {et~al.}(2021)\citenamefont {Xie},
  \citenamefont {Wang}, \citenamefont {Zhang}, \citenamefont {Zhan},
  \citenamefont {Jiang}, \citenamefont {Lu},\ and\ \citenamefont
  {Chen}}]{xie_higher-order_2021}%
  \BibitemOpen
  \bibfield  {author} {\bibinfo {author} {\bibfnamefont {B.}~\bibnamefont
  {Xie}}, \bibinfo {author} {\bibfnamefont {H.-X.}\ \bibnamefont {Wang}},
  \bibinfo {author} {\bibfnamefont {X.}~\bibnamefont {Zhang}}, \bibinfo
  {author} {\bibfnamefont {P.}~\bibnamefont {Zhan}}, \bibinfo {author}
  {\bibfnamefont {J.-H.}\ \bibnamefont {Jiang}}, \bibinfo {author}
  {\bibfnamefont {M.}~\bibnamefont {Lu}}, \ and\ \bibinfo {author}
  {\bibfnamefont {Y.}~\bibnamefont {Chen}},\ }\href {\doibase
  10.1038/s42254-021-00323-4} {\bibfield  {journal} {\bibinfo  {journal} {Nat
  Rev Phys}\ }\textbf {\bibinfo {volume} {3}},\ \bibinfo {pages} {520}
  (\bibinfo {year} {2021})}\BibitemShut {NoStop}%
\bibitem [{\citenamefont {Yang}\ \emph {et~al.}(2024)\citenamefont {Yang},
  \citenamefont {Wang}, \citenamefont {Li},\ and\ \citenamefont
  {Xu}}]{yang_higher-order_2024}%
  \BibitemOpen
  \bibfield  {author} {\bibinfo {author} {\bibfnamefont {Y.-B.}\ \bibnamefont
  {Yang}}, \bibinfo {author} {\bibfnamefont {J.-H.}\ \bibnamefont {Wang}},
  \bibinfo {author} {\bibfnamefont {K.}~\bibnamefont {Li}}, \ and\ \bibinfo
  {author} {\bibfnamefont {Y.}~\bibnamefont {Xu}},\ }\href {\doibase
  10.1088/1361-648X/ad3abd} {\bibfield  {journal} {\bibinfo  {journal} {J.
  Phys.: Condens. Matter}\ }\textbf {\bibinfo {volume} {36}},\ \bibinfo {pages}
  {283002} (\bibinfo {year} {2024})}\BibitemShut {NoStop}%
\bibitem [{\citenamefont {Kang}\ \emph {et~al.}(2019)\citenamefont {Kang},
  \citenamefont {Shiozaki},\ and\ \citenamefont {Cho}}]{kang_many-body_2019}%
  \BibitemOpen
  \bibfield  {author} {\bibinfo {author} {\bibfnamefont {B.}~\bibnamefont
  {Kang}}, \bibinfo {author} {\bibfnamefont {K.}~\bibnamefont {Shiozaki}}, \
  and\ \bibinfo {author} {\bibfnamefont {G.~Y.}\ \bibnamefont {Cho}},\ }\href
  {\doibase 10.1103/PhysRevB.100.245134} {\bibfield  {journal} {\bibinfo
  {journal} {Phys. Rev. B}\ }\textbf {\bibinfo {volume} {100}},\ \bibinfo
  {pages} {245134} (\bibinfo {year} {2019})}\BibitemShut {NoStop}%
\bibitem [{\citenamefont {Ono}\ \emph {et~al.}(2019)\citenamefont {Ono},
  \citenamefont {Trifunovic},\ and\ \citenamefont
  {Watanabe}}]{ono_difficulties_2019}%
  \BibitemOpen
  \bibfield  {author} {\bibinfo {author} {\bibfnamefont {S.}~\bibnamefont
  {Ono}}, \bibinfo {author} {\bibfnamefont {L.}~\bibnamefont {Trifunovic}}, \
  and\ \bibinfo {author} {\bibfnamefont {H.}~\bibnamefont {Watanabe}},\ }\href
  {\doibase 10.1103/PhysRevB.100.245133} {\bibfield  {journal} {\bibinfo
  {journal} {Phys. Rev. B}\ }\textbf {\bibinfo {volume} {100}},\ \bibinfo
  {pages} {245133} (\bibinfo {year} {2019})}\BibitemShut {NoStop}%
\bibitem [{\citenamefont {Lee}\ \emph {et~al.}(2022)\citenamefont {Lee},
  \citenamefont {Cho},\ and\ \citenamefont {Kang}}]{lee_many-body_2022}%
  \BibitemOpen
  \bibfield  {author} {\bibinfo {author} {\bibfnamefont {W.}~\bibnamefont
  {Lee}}, \bibinfo {author} {\bibfnamefont {G.~Y.}\ \bibnamefont {Cho}}, \ and\
  \bibinfo {author} {\bibfnamefont {B.}~\bibnamefont {Kang}},\ }\href {\doibase
  10.1103/PhysRevB.105.155143} {\bibfield  {journal} {\bibinfo  {journal}
  {Phys. Rev. B}\ }\textbf {\bibinfo {volume} {105}},\ \bibinfo {pages}
  {155143} (\bibinfo {year} {2022})}\BibitemShut {NoStop}%
\bibitem [{\citenamefont {Benalcazar}\ \emph
  {et~al.}(2017{\natexlab{b}})\citenamefont {Benalcazar}, \citenamefont
  {Bernevig},\ and\ \citenamefont {Hughes}}]{benalcazar_electric_2017}%
  \BibitemOpen
  \bibfield  {author} {\bibinfo {author} {\bibfnamefont {W.~A.}\ \bibnamefont
  {Benalcazar}}, \bibinfo {author} {\bibfnamefont {B.~A.}\ \bibnamefont
  {Bernevig}}, \ and\ \bibinfo {author} {\bibfnamefont {T.~L.}\ \bibnamefont
  {Hughes}},\ }\href {\doibase 10.1103/PhysRevB.96.245115} {\bibfield
  {journal} {\bibinfo  {journal} {Phys. Rev. B}\ }\textbf {\bibinfo {volume}
  {96}},\ \bibinfo {pages} {245115} (\bibinfo {year}
  {2017}{\natexlab{b}})}\BibitemShut {NoStop}%
\bibitem [{\citenamefont {Nourafkan}\ and\ \citenamefont
  {Kotliar}(2013)}]{nourafkan_electric_2013}%
  \BibitemOpen
  \bibfield  {author} {\bibinfo {author} {\bibfnamefont {R.}~\bibnamefont
  {Nourafkan}}\ and\ \bibinfo {author} {\bibfnamefont {G.}~\bibnamefont
  {Kotliar}},\ }\href {\doibase 10.1103/PhysRevB.88.155121} {\bibfield
  {journal} {\bibinfo  {journal} {Phys. Rev. B}\ }\textbf {\bibinfo {volume}
  {88}},\ \bibinfo {pages} {155121} (\bibinfo {year} {2013})}\BibitemShut
  {NoStop}%
\bibitem [{\citenamefont {Georges}\ \emph {et~al.}(1996)\citenamefont
  {Georges}, \citenamefont {Kotliar}, \citenamefont {Krauth},\ and\
  \citenamefont {Rozenberg}}]{georges_dynamical_1996}%
  \BibitemOpen
  \bibfield  {author} {\bibinfo {author} {\bibfnamefont {A.}~\bibnamefont
  {Georges}}, \bibinfo {author} {\bibfnamefont {G.}~\bibnamefont {Kotliar}},
  \bibinfo {author} {\bibfnamefont {W.}~\bibnamefont {Krauth}}, \ and\ \bibinfo
  {author} {\bibfnamefont {M.~J.}\ \bibnamefont {Rozenberg}},\ }\href {\doibase
  10.1103/RevModPhys.68.13} {\bibfield  {journal} {\bibinfo  {journal} {Rev.
  Mod. Phys.}\ }\textbf {\bibinfo {volume} {68}},\ \bibinfo {pages} {13}
  (\bibinfo {year} {1996})}\BibitemShut {NoStop}%
\bibitem [{\citenamefont {Peng}\ \emph {et~al.}(2020)\citenamefont {Peng},
  \citenamefont {He},\ and\ \citenamefont {Lu}}]{peng_correlation_2020}%
  \BibitemOpen
  \bibfield  {author} {\bibinfo {author} {\bibfnamefont {C.}~\bibnamefont
  {Peng}}, \bibinfo {author} {\bibfnamefont {R.-Q.}\ \bibnamefont {He}}, \ and\
  \bibinfo {author} {\bibfnamefont {Z.-Y.}\ \bibnamefont {Lu}},\ }\href
  {\doibase 10.1103/PhysRevB.102.045110} {\bibfield  {journal} {\bibinfo
  {journal} {Phys. Rev. B}\ }\textbf {\bibinfo {volume} {102}},\ \bibinfo
  {pages} {045110} (\bibinfo {year} {2020})},\ \bibinfo {note} {publisher:
  American Physical Society}\BibitemShut {NoStop}%
\bibitem [{\citenamefont {Kang}\ \emph {et~al.}(2021)\citenamefont {Kang},
  \citenamefont {Lee},\ and\ \citenamefont {Cho}}]{kang_many-body_2021}%
  \BibitemOpen
  \bibfield  {author} {\bibinfo {author} {\bibfnamefont {B.}~\bibnamefont
  {Kang}}, \bibinfo {author} {\bibfnamefont {W.}~\bibnamefont {Lee}}, \ and\
  \bibinfo {author} {\bibfnamefont {G.~Y.}\ \bibnamefont {Cho}},\ }\href
  {\doibase 10.1103/PhysRevLett.126.016402} {\bibfield  {journal} {\bibinfo
  {journal} {Phys. Rev. Lett.}\ }\textbf {\bibinfo {volume} {126}},\ \bibinfo
  {pages} {016402} (\bibinfo {year} {2021})}\BibitemShut {NoStop}%
\bibitem [{\citenamefont
  {Potthoff}(2003)}]{potthoff_self-energy-functional_2003}%
  \BibitemOpen
  \bibfield  {author} {\bibinfo {author} {\bibfnamefont {M.}~\bibnamefont
  {Potthoff}},\ }\href {\doibase 10.1140/epjb/e2003-00121-8} {\bibfield
  {journal} {\bibinfo  {journal} {Eur. Phys. J. B}\ }\textbf {\bibinfo {volume}
  {32}},\ \bibinfo {pages} {429} (\bibinfo {year} {2003})}\BibitemShut
  {NoStop}%
\bibitem [{\citenamefont {Helmes}(2008)}]{helmes_dynamical_2008}%
  \BibitemOpen
  \bibfield  {author} {\bibinfo {author} {\bibfnamefont {R.}~\bibnamefont
  {Helmes}},\ }\emph {\bibinfo {title} {Dynamical {Mean} {Field} {Thoery} of
  inhomogeneous correlated systems}},\ \href@noop {} {\bibinfo {type} {{PhD}
  {Thesis}}},\ \bibinfo  {school} {Koln University} (\bibinfo {year}
  {2008})\BibitemShut {NoStop}%
\bibitem [{\citenamefont {Hofstetter}\ and\ \citenamefont
  {Qin}(2018)}]{hofstetter_quantum_2018}%
  \BibitemOpen
  \bibfield  {author} {\bibinfo {author} {\bibfnamefont {W.}~\bibnamefont
  {Hofstetter}}\ and\ \bibinfo {author} {\bibfnamefont {T.}~\bibnamefont
  {Qin}},\ }\href {http://stacks.iop.org/0953-4075/51/i=8/a=082001} {\bibfield
  {journal} {\bibinfo  {journal} {Journal of Physics B: Atomic, Molecular and
  Optical Physics}\ }\textbf {\bibinfo {volume} {51}},\ \bibinfo {pages}
  {082001} (\bibinfo {year} {2018})}\BibitemShut {NoStop}%
\bibitem [{\citenamefont {Resta}(2006)}]{resta_polarization_2006}%
  \BibitemOpen
  \bibfield  {author} {\bibinfo {author} {\bibfnamefont {R.}~\bibnamefont
  {Resta}},\ }\href {\doibase 10.1103/PhysRevLett.96.137601} {\bibfield
  {journal} {\bibinfo  {journal} {Phys. Rev. Lett.}\ }\textbf {\bibinfo
  {volume} {96}},\ \bibinfo {pages} {137601} (\bibinfo {year}
  {2006})}\BibitemShut {NoStop}%
\bibitem [{\citenamefont {Souza}\ \emph {et~al.}(2000)\citenamefont {Souza},
  \citenamefont {Wilkens},\ and\ \citenamefont
  {Martin}}]{souza_polarization_2000}%
  \BibitemOpen
  \bibfield  {author} {\bibinfo {author} {\bibfnamefont {I.}~\bibnamefont
  {Souza}}, \bibinfo {author} {\bibfnamefont {T.}~\bibnamefont {Wilkens}}, \
  and\ \bibinfo {author} {\bibfnamefont {R.~M.}\ \bibnamefont {Martin}},\
  }\href {\doibase 10.1103/PhysRevB.62.1666} {\bibfield  {journal} {\bibinfo
  {journal} {Phys. Rev. B}\ }\textbf {\bibinfo {volume} {62}},\ \bibinfo
  {pages} {1666} (\bibinfo {year} {2000})}\BibitemShut {NoStop}%
\bibitem [{\citenamefont {Cheong}\ and\ \citenamefont
  {Henley}(2004)}]{cheong_many-body_2004}%
  \BibitemOpen
  \bibfield  {author} {\bibinfo {author} {\bibfnamefont {S.-A.}\ \bibnamefont
  {Cheong}}\ and\ \bibinfo {author} {\bibfnamefont {C.~L.}\ \bibnamefont
  {Henley}},\ }\href {\doibase 10.1103/PhysRevB.69.075111} {\bibfield
  {journal} {\bibinfo  {journal} {Phys. Rev. B}\ }\textbf {\bibinfo {volume}
  {69}},\ \bibinfo {pages} {075111} (\bibinfo {year} {2004})}\BibitemShut
  {NoStop}%
\bibitem [{\citenamefont {Meyer}(2023)}]{meyer_matrix_2023}%
  \BibitemOpen
  \bibfield  {author} {\bibinfo {author} {\bibfnamefont {C.~D.}\ \bibnamefont
  {Meyer}},\ }\href {\doibase 10.1137/1.9781611977448} {\emph {\bibinfo {title}
  {Matrix {Analysis} and {Applied} {Linear} {Algebra}, {Second} {Edition}}}}\
  (\bibinfo  {publisher} {Society for Industrial and Applied Mathematics},\
  \bibinfo {address} {Philadelphia, PA},\ \bibinfo {year} {2023})\BibitemShut
  {NoStop}%
\bibitem [{\citenamefont {Resta}(2021)}]{resta_dipole_2021}%
  \BibitemOpen
  \bibfield  {author} {\bibinfo {author} {\bibfnamefont {R.}~\bibnamefont
  {Resta}},\ }\href {\doibase 10.1063/5.0040815} {\bibfield  {journal}
  {\bibinfo  {journal} {J. Chem. Phys.}\ }\textbf {\bibinfo {volume} {154}},\
  \bibinfo {pages} {050901} (\bibinfo {year} {2021})}\BibitemShut {NoStop}%
\bibitem [{\citenamefont {Tran}\ \emph {et~al.}(2015)\citenamefont {Tran},
  \citenamefont {Dauphin}, \citenamefont {Goldman},\ and\ \citenamefont
  {Gaspard}}]{tran_topological_2015}%
  \BibitemOpen
  \bibfield  {author} {\bibinfo {author} {\bibfnamefont {D.-T.}\ \bibnamefont
  {Tran}}, \bibinfo {author} {\bibfnamefont {A.}~\bibnamefont {Dauphin}},
  \bibinfo {author} {\bibfnamefont {N.}~\bibnamefont {Goldman}}, \ and\
  \bibinfo {author} {\bibfnamefont {P.}~\bibnamefont {Gaspard}},\ }\href
  {\doibase 10.1103/PhysRevB.91.085125} {\bibfield  {journal} {\bibinfo
  {journal} {Phys. Rev. B}\ }\textbf {\bibinfo {volume} {91}},\ \bibinfo
  {pages} {085125} (\bibinfo {year} {2015})}\BibitemShut {NoStop}%
\bibitem [{\citenamefont {Agarwala}\ and\ \citenamefont
  {Shenoy}(2017)}]{agarwala_topological_2017}%
  \BibitemOpen
  \bibfield  {author} {\bibinfo {author} {\bibfnamefont {A.}~\bibnamefont
  {Agarwala}}\ and\ \bibinfo {author} {\bibfnamefont {V.~B.}\ \bibnamefont
  {Shenoy}},\ }\href {\doibase 10.1103/PhysRevLett.118.236402} {\bibfield
  {journal} {\bibinfo  {journal} {Phys. Rev. Lett.}\ }\textbf {\bibinfo
  {volume} {118}},\ \bibinfo {pages} {236402} (\bibinfo {year} {2017})},\
  \bibinfo {note} {publisher: American Physical Society}\BibitemShut {NoStop}%
\bibitem [{\citenamefont {Yang}\ \emph {et~al.}(2019)\citenamefont {Yang},
  \citenamefont {Qin}, \citenamefont {Deng}, \citenamefont {Duan},\ and\
  \citenamefont {Xu}}]{yang_topological_2019}%
  \BibitemOpen
  \bibfield  {author} {\bibinfo {author} {\bibfnamefont {Y.-B.}\ \bibnamefont
  {Yang}}, \bibinfo {author} {\bibfnamefont {T.}~\bibnamefont {Qin}}, \bibinfo
  {author} {\bibfnamefont {D.-L.}\ \bibnamefont {Deng}}, \bibinfo {author}
  {\bibfnamefont {L.-M.}\ \bibnamefont {Duan}}, \ and\ \bibinfo {author}
  {\bibfnamefont {Y.}~\bibnamefont {Xu}},\ }\href {\doibase
  10.1103/PhysRevLett.123.076401} {\bibfield  {journal} {\bibinfo  {journal}
  {Phys. Rev. Lett.}\ }\textbf {\bibinfo {volume} {123}},\ \bibinfo {pages}
  {076401} (\bibinfo {year} {2019})}\BibitemShut {NoStop}%
\bibitem [{\citenamefont {Vidberg}\ and\ \citenamefont
  {Serene}(1977)}]{vidberg_solving_1977}%
  \BibitemOpen
  \bibfield  {author} {\bibinfo {author} {\bibfnamefont {H.~J.}\ \bibnamefont
  {Vidberg}}\ and\ \bibinfo {author} {\bibfnamefont {J.~W.}\ \bibnamefont
  {Serene}},\ }\href {\doibase 10.1007/BF00655090} {\bibfield  {journal}
  {\bibinfo  {journal} {Journal of Low Temperature Physics}\ }\textbf {\bibinfo
  {volume} {29}},\ \bibinfo {pages} {179} (\bibinfo {year} {1977})}\BibitemShut
  {NoStop}%
\bibitem [{\citenamefont {Su}\ \emph {et~al.}(1980)\citenamefont {Su},
  \citenamefont {Schrieffer},\ and\ \citenamefont {Heeger}}]{su_soliton_1980}%
  \BibitemOpen
  \bibfield  {author} {\bibinfo {author} {\bibfnamefont {W.~P.}\ \bibnamefont
  {Su}}, \bibinfo {author} {\bibfnamefont {J.~R.}\ \bibnamefont {Schrieffer}},
  \ and\ \bibinfo {author} {\bibfnamefont {A.~J.}\ \bibnamefont {Heeger}},\
  }\href {\doibase 10.1103/PhysRevB.22.2099} {\bibfield  {journal} {\bibinfo
  {journal} {Phys. Rev. B}\ }\textbf {\bibinfo {volume} {22}},\ \bibinfo
  {pages} {2099} (\bibinfo {year} {1980})}\BibitemShut {NoStop}%
\bibitem [{\citenamefont {Yatsugi}\ \emph {et~al.}(2022)\citenamefont
  {Yatsugi}, \citenamefont {Yoshida}, \citenamefont {Mizoguchi}, \citenamefont
  {Kuno}, \citenamefont {Iizuka}, \citenamefont {Tadokoro},\ and\ \citenamefont
  {Hatsugai}}]{yatsugi_observation_2022}%
  \BibitemOpen
  \bibfield  {author} {\bibinfo {author} {\bibfnamefont {K.}~\bibnamefont
  {Yatsugi}}, \bibinfo {author} {\bibfnamefont {T.}~\bibnamefont {Yoshida}},
  \bibinfo {author} {\bibfnamefont {T.}~\bibnamefont {Mizoguchi}}, \bibinfo
  {author} {\bibfnamefont {Y.}~\bibnamefont {Kuno}}, \bibinfo {author}
  {\bibfnamefont {H.}~\bibnamefont {Iizuka}}, \bibinfo {author} {\bibfnamefont
  {Y.}~\bibnamefont {Tadokoro}}, \ and\ \bibinfo {author} {\bibfnamefont
  {Y.}~\bibnamefont {Hatsugai}},\ }\href {\doibase 10.1038/s42005-022-00957-5}
  {\bibfield  {journal} {\bibinfo  {journal} {Commun Phys}\ }\textbf {\bibinfo
  {volume} {5}},\ \bibinfo {pages} {180} (\bibinfo {year} {2022})}\BibitemShut
  {NoStop}%
\bibitem [{\citenamefont {Dong}\ \emph {et~al.}(2021)\citenamefont {Dong},
  \citenamefont {Juri\v{c}i\'{c}},\ and\ \citenamefont
  {Roy}}]{dong_topolectric_2021-1}%
  \BibitemOpen
  \bibfield  {author} {\bibinfo {author} {\bibfnamefont {J.}~\bibnamefont
  {Dong}}, \bibinfo {author} {\bibfnamefont {V.}~\bibnamefont
  {Juri\v{c}i\'{c}}}, \ and\ \bibinfo {author} {\bibfnamefont {B.}~\bibnamefont
  {Roy}},\ }\href {\doibase 10.1103/PhysRevResearch.3.023056} {\bibfield
  {journal} {\bibinfo  {journal} {Phys. Rev. Res.}\ }\textbf {\bibinfo {volume}
  {3}},\ \bibinfo {pages} {023056} (\bibinfo {year} {2021})},\ \bibinfo {note}
  {publisher: American Physical Society}\BibitemShut {NoStop}%
\end{thebibliography}
%

\end{document}